\documentclass[12pt,preprint]{aastex}
\usepackage{natbib}

\newcommand{\asca}{{\small \it ASCA}}

\newcommand{\rosat}{{\small \it ROSAT}}

\newcommand{\ginga}{{\it Ginga}}
\newcommand{\sax}{{\small \it BeppoSA$\!$X}}
\newcommand{\xte}{{\small \it XTE}}

\newcommand{\agn}{{\small AGN}}
\newcommand{\xstar}{{\small XSTAR}}

\newcommand{\mcg}{MCG$-$6-30-15}
\newcommand{\mrk}{Mrk~766}
\newcommand{\iras}{IRAS~13349+2438}
\newcommand{\xmm}{{\it XMM-Newton}}
\newcommand{\rgs}{{\small RGS}}
\newcommand{\epic}{{\small EPIC}}
\newcommand{\ccd}{{\small CCD}}
\newcommand{\mos}{{\small MOS}}
\newcommand{\pn}{{\small PN}}
\newcommand{\om}{{\small OM}}
\newcommand{\chandra}{{\it Chandra}}
\newcommand{\hetg}{{\small HETG}}
\newcommand{\hullac}{{\small HULLAC}}
\newcommand{\uta}{{\small UTA}}
\newcommand{\ew}{{\small EW}}
\newcommand{\kev}{ke~\hspace{-0.18cm}V}
\newcommand{\ev}{e~\hspace{-0.18cm}V}
\newcommand{\nasa}{{\small NASA}}
\newcommand{\esa}{{\small ESA}}

\begin{document}

\title{Can a Dusty Warm Absorber Model Reproduce the Soft X-ray Spectra of
       \mcg\ and \mrk?}
\author{Masao Sako\altaffilmark{1,2,3}, Steven M. Kahn\altaffilmark{1},
        Graziella Branduardi-Raymont\altaffilmark{4}, Jelle S.
        Kaastra\altaffilmark{5},Albert C. Brinkman\altaffilmark{5}, 
        Mathew J. Page\altaffilmark{4}, Ehud Behar\altaffilmark{1},
        Frits Paerels\altaffilmark{1}, Ali Kinkhabwala\altaffilmark{1},
        Duane A. Liedahl\altaffilmark{6},
        \& Jan Willem den Herder\altaffilmark{5}}

\altaffiltext{1}{Columbia Astrophysics Laboratory,
                 550 West 120th Street, New York, NY 10027, USA;
                 skahn@astro.columbia.edu,
                 behar@astro.columbia.edu,
                 frits@astro.columbia.edu,
                 ali@astro.columbia.edu}
\altaffiltext{2}{Present address:  Theoretical Astrophysics and
                 Space Radiation Laboratory,
                 California Institute of Technology,
                 MC 130-33, Pasadena, CA 91125, USA;
                 masao@tapir.caltech.edu}
\altaffiltext{3}{Chandra Postdoctoral Fellow}
\altaffiltext{4}{Mullard Space Science Laboratory,
                 University College London,
                 Holmbury St. Mary, Dorking, Surrey,
                 RH5 6NT, UK; gbr@mssl.ucl.ac.uk,
                 mjp@mssl.ucl.ac.uk}
\altaffiltext{5}{SRON, the National Institute for Space Research,
                 Sorbonnelaan 2, 3584 CA Utrecht, The Netherlands;
                 J.S.Kaastra@sron.nl, A.C.Brinkman@sron.nl,
                 J.W.A.den.Herder@sronl.nl}
\altaffiltext{6}{Physics Department,
                 Lawrence Livermore National Laboratory,
                 P.O. Box 808, L-41, Livermore, CA 94550, USA;
                 duane@virgo.llnl.gov}


\slugcomment{Accepted for Publication to ApJ}
\shorttitle{Soft X-ray Spectra of \mcg\ and \mrk}
\shortauthors{Sako et al.}

\begin{abstract}

  \xmm\ \rgs\ spectra of \mcg\ and \mrk\ exhibit complex discrete structure,
  which was interpreted in a paper by \citet{grazie01} as evidence for the
  existence of relativistically broadened Lyman alpha emission from carbon,
  nitrogen, and oxygen, produced in the inner-most regions of an accretion
  disk around a Kerr black hole.  This suggestion was subsequently criticized
  in a paper by \citet{lee01}, who argued that for \mcg, the \chandra\ \hetg\
  spectrum, which is partially overlapping the \rgs\ in spectral coverage, is
  adequately fit by a dusty warm absorber model, with no relativistic line
  emission.  We present a reanalysis of the original \rgs\ data sets in terms
  of the \citet{lee01} model.  Specifically, we show that: (1) The explicit
  model given by \citet{lee01} differs markedly from the \rgs\ data,
  especially at longer wavelengths, beyond the region sampled by the \hetg;
  (2) Generalizations of the \citet{lee01} model, with all parameters left
  free, do provide qualitatively better fits to the \rgs\ data, but are still
  incompatible with the detailed spectral structure; (3) The ionized oxygen
  absorption line equivalent widths are well-measured with the \rgs\ for both
  sources, and place very tight constraints on both the column densities and
  turbulent velocity widths of \ion{O}{7} and \ion{O}{8}.  The derived column
  densities are well below those posited by \citet{lee01}, and are
  insufficient to play any role in explaining the observed edge-like feature
  near $17.5$~\AA; (4) The lack of a significant neutral oxygen edge near
  $23$~\AA\ places very strong limits on any possible contribution of
  absorption to the observed structure by dust embedded in a warm medium; (5)
  The original relativistic line model with warm absorption proposed by
  \citet{grazie01} provides a superior fit to the \rgs\ data, both in the
  overall shape of the spectrum and in the discrete absorption lines.  We also
  discuss a possible theoretical interpretation for the putative relativistic
  Lyman alpha line emission in terms of the photoionized surface layers of the
  inner regions of an accretion disk.  While there are still a number of
  outstanding theoretical questions about the viability of such a model, it is
  interesting to note that simple estimates of key parameters are roughly
  compatible with those derived from the observed spectra.

\end{abstract}


\section{Introduction}
  High-resolution soft X-ray spectra of Seyfert~1 galaxies acquired with the
  grating spectrometers onboard the \chandra\ and \xmm\ observatories have
  allowed us, for the first time, to investigate both the dynamics of the
  extended absorbing media in these systems, and the structure of the
  intrinsic spectra that originate in the vicinity of the central massive
  black holes.  To date, narrow X-ray absorption lines have been detected in
  many objects, including NGC~5548 \citep{kaastra00}, NGC~3783
  \citep{kaspi00,kaspi01}, \mcg\ and \mrk\ \citep{grazie01,lee01},
  IRAS~13349+2438 \citep{sako01}, NGC~4051 \citep{collinge01}, and Mrk~509
  \citep{pounds01}.  On the other hand, the spectra of several sources show no
  evidence for absorption, but, instead, exhibit complex spectral features and
  temporal behavior that cannot be explained by simple continuum emission
  models, and are more likely to be related to the activity of the central
  engine (e.g., Ton~S180, \citealt{turnertj01}; NGC~4593,
  \citealt{katrien01}).

  In an earlier paper presenting high-resolution spectra of \mcg\ and \mrk\
  obtained with the Reflection Grating Spectrometer (\rgs) onboard \xmm\
  \citep{grazie01}, we suggested that the soft X-ray emission of these two
  objects in the $\lambda = 5 - 38$~\AA\ ($E = 0.35 - 2.5 ~\rm{\kev}$) band
  includes significant contributions from discrete emission features produced
  through X-ray illumination of an accretion disk around the central black
  hole.  The observed spectral structure matches well with what is expected
  for relativistically broadened recombination lines of hydrogen-like carbon,
  nitrogen, and oxygen originating from the inner regions of the disk in a
  Kerr metric.  That interpretation significantly challenges conventional
  models of the global soft X-ray spectra of \agn, where it is usually assumed
  that all discrete structure in the soft X-ray band is due to absorption
  features produced in an extended medium far away from the central black
  hole.

  In a subsequent paper, \citet{lee01} questioned our interpretation based on
  their analysis of a non-simultaneous \chandra\ \hetg\ observation of \mcg.
  They claimed that the $\lambda < 25$~\AA\ ($E > 0.5 ~\rm{\kev}$) region can
  be well-reproduced solely by a dusty warm absorber model superimposed on a
  smooth continuum, and does not require the presence of any relativistic disk
  emission lines.  They interpret the excess soft X-ray emission above the
  high energy power law continuum as either a blackbody component, or a
  steepening of the power law in the soft X-ray band.

  The purpose of this paper is to provide a reanalysis of the \xmm\ data
  presented by \citet{grazie01}, with more detailed attention to the
  properties of the discrete absorption components in both sources.  For \mcg,
  we explicitly test the spectral model proposed by \citet{lee01}, and
  demonstrate that, contrary to their claim, it cannot account for all of the
  features observed in the \rgs\ spectrum.  In particular, the absorption
  model fails to self-consistently fit the details of both the oxygen line
  absorption and the apparent edge-like structure present in the data.
  Further, the \citet{lee01} hypothesis that much of the inferred absorption
  can be ascribed to warm dust, is clearly ruled out by the \rgs\ spectra.  We
  reconfirm our earlier conclusion that the global spectrum strongly favors
  the intrinsic nuclear radiation to be dominated by discrete emission line
  features that are expected to be formed in an X-ray photoionized accretion
  disk.  The observed line profiles indicate that they are distorted through
  relativistic beaming and smeared by strong gravitational effects in the
  vicinity of the black hole.

  This paper is organized as follows.  In \S2, we summarize the observations
  and describe the procedures adopted for data reduction.  While we focus
  primarily on the spectra obtained by the \rgs, we adopt continuum parameters
  derived from the \epic\ data in the $2 - 10 ~\rm{\kev}$ region, which is
  mostly outside the \rgs\ bandpass.  The procedures and results are
  summarized in \S3.  In \S4, we discuss the model proposed by \citet{lee01}
  and compare it directly to the \rgs\ data.  We also describe in detail our
  measurements of the ionized absorber parameters, as well as the possible
  existence of absorption by dust and the limits on the column densities.
  Having characterized the contributions of the extended absorbing medium, we
  demonstrate that the intrinsic spectra of both \mcg\ and \mrk\ are not
  smooth, but instead consist of discrete jumps.  We interpret these jumps as
  blue emission edges of carbon, nitrogen, and oxygen Ly$\alpha$ recombination
  lines produced in the inner regions of a relativistic accretion disk, which
  are described in \S5.  We briefly summarize our results in \S6.

\section{Observations and Data Reduction}
  \mrk\ was observed with \xmm\ \citep{jansen01} during the Performance
  Verification phase on May 20, 2000.  \mcg\ was observed during the
  Guaranteed Time Observations phase on July $10~-~11$, 2000.  The instrument
  modes for the Reflection Grating Spectrometer (\rgs; \citealt{denherder01}),
  the European Photon Imaging Cameras (\epic) \mos\ (\citealt{turner01}) and
  \pn\ (\citealt{struder01}), and the Optical Monitor (\om;
  \citealt{mason01}), as well as the associated exposure times for both
  observations, are summarized in Table~\ref{tbl1}.

  The data were processed through the \xmm\ Science Analysis Software ({\small
  SAS} v5.3.3) using the calibration database released with the software.  For
  the \rgs, we extract events from a region in dispersion/cross-dispersion
  coordinates, which contains 90\% of the source counts.  The data are then
  screened through a dispersion/pulse-height filter to extract the first order
  dispersed photons.  This region contains 90\% of the \ccd\ pulse-height
  redistribution function.  Dispersion coordinates are then converted to
  photon wavelength using the most up-to-date calibration data.  The resulting
  wavelengths are accurate to within $\sim 8$~m\AA\ \citep{denherder01}, which
  corresponds to a velocity uncertainty of $\sim 120 ~\rm{km~s}^{-1}$ at
  20~\AA.  The background spectrum was generated using the entire region of
  the \ccd\ array excluded by a 95\% source extraction region.  The effective
  areas were calculated for the extraction regions described above.  The data
  used here are identical to those presented in \citet{grazie01}.  The
  effective areas, however, are slightly updated to correctly account the
  shape and depth of the instrumental oxygen edge particlularly near
  threshold.  In addition, chip-to-chip variations in the \ccd\ quantum
  efficiencies ($\sim 10$\%) have been calibrated using deep exposures of
  several bright continuum sources.  Both of these features have been
  incorporated into the calibration database released with {\small SAS}
  v5.3.3.

  We use the time-integrated spectrum for all of our spectral analyses.  In
  all cases, \rgs1 and \rgs2 data are simutaneously fit, although in most
  cases, we show only the \rgs1 for clarity.  For both \mcg\ and \mrk, the
  spectrum in the wavelength range $20.0 - 24.0$~\AA\ is not covered by \rgs2
  due to a malfunctioning of \ccd4.  The observations were, however, performed
  prior to the malfunctioning of \ccd7 on \rgs1, which cover the range
  $\lambda = 10.5 - 13.7$~\AA.  The \rgs\ spectrum of \mcg\ is of high
  statistical quality with $\sim 190 ~\rm{counts}$ per $15$ m\AA\ bin at the
  peak of the spectrum near $\lambda \sim 18 - 20 $~\AA, which is $\ga 10$
  times higher than that of the \chandra\ \hetg\ data presented in
  \citet{lee01} in the same spectral region.  Near the \ion{O}{7} He$\alpha$
  complex ($\lambda = 22$\AA), which is not covered by \rgs2, the \rgs1
  spectrum contains only $\sim 75 ~\rm{counts}$ per $15$ m\AA\ bin.  This is
  still $\sim 8$ more than in the \chandra\ spectrum, since the \hetg\
  efficiency drops by a factor of $\sim 2$ from 19~\AA\ to 22~\AA, while the
  \rgs\ effective area decreases by only $\sim 20$\%.  The resolving power of
  the \rgs\, however, is a factor of $\sim 3$ lower, so the figure of merit
  for detecting isolated lines, which is proportional to
  $(A_{\rm{eff}}R)^{1/2}$ where $A_{\rm{eff}}$ is the effective area and $R$
  is the resolving power, is $1.6 - 1.8$ times higher with the \rgs\ for
  $\lambda \ga 18$~\AA.  The spectrum of \mrk\ is substantially noisier, with
  $\sim 70 ~\rm{counts}$ per $15$ m\AA\ bin in the same region.

  The spectra of \mcg\ used for fitting are binned using a constant dispersion
  width of $1.2 \times 10^{-5} ~\rm{radians}$, which corresponds to
  approximately $0.017$\AA\ at $10$\AA\ and $0.024$\AA\ at $25$~\AA.
  Throughout most of the bandpass ($7 - 30$\AA), each spectral bin contains at
  least 20 counts.  The bin sizes adopted for the \mrk\ spectra are larger by
  a factor of 2 owing to the fewer number of counts detected.  Unless
  otherwise stated, the data are also plotted using these binning schemes.

  The \epic\ \mos\ and \pn\ spectra were generated from a circular extraction
  region centered on the source with radii of $1\arcmin$ and $42\arcsec$ for
  \mcg\ and \mrk, respectively.  The effective area curves were calculated
  based on this extraction region.  Background spectra were generated using
  events that lie on off-axis regions on the same chip.  For the purposes of
  our analyses, we use the \epic\ data only for the determination of the high
  energy power law continuum, but as shown in \citet{page01} in the case of
  \mrk, the general characteristics of the spectra from all three instruments
  (\rgs, \epic\ \pn\ and \mos\ ) agree well to within the instruments'
  calibration uncertainties.  The same is true for \mcg.

  Throughout the analysis, we adopt cosmologial redshifts of $z = 0.01293$ for
  \mrk\ \citep{smith87} and $z = 0.00775$ for \mcg\ \citep{fisher95}, and
  Galactic absorption column densities of $N_{\rm H} = 4.1 \times 10^{20}
  ~\rm{cm}^{-2}$ for \mcg\ \citep{elvis89} and $N_{\rm H} = 1.8 \times 10^{20}
  ~\rm{cm}^{-2}$ for \mrk\ \citep{murphy96}.  The uncertainties in the
  redshifts ($\Delta z = 0.00005$) are much smaller than the resolving power
  capabilities of the instrument.  All errors quoted throughout the paper
  correspond to 90\% uncertainties for a single parameter unless otherwise
  stated.

\section{High Energy Spectrum}
\begin{figure}
  \includegraphics[height=4in,angle=-90]{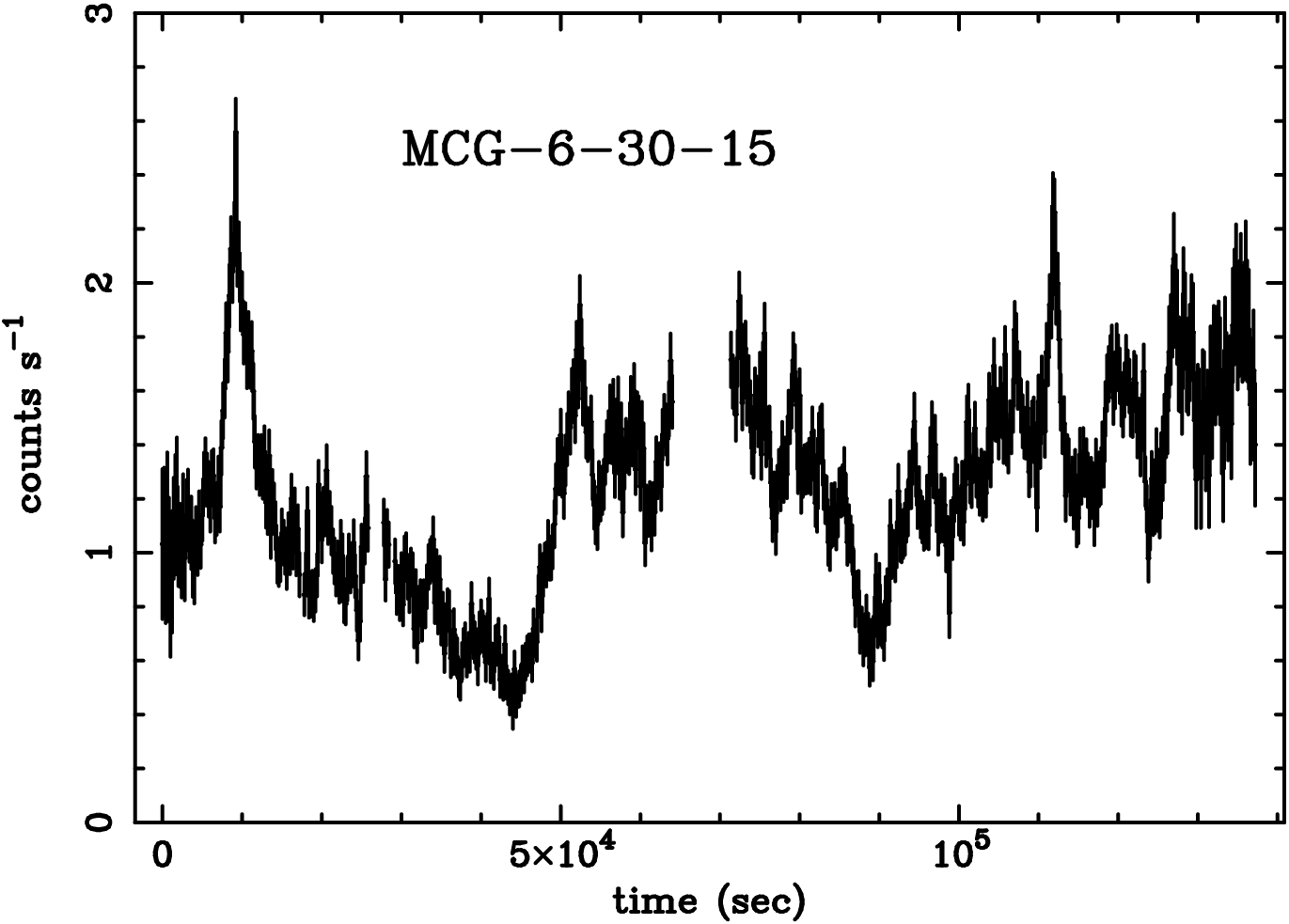}\\
  \includegraphics[height=4in,angle=-90]{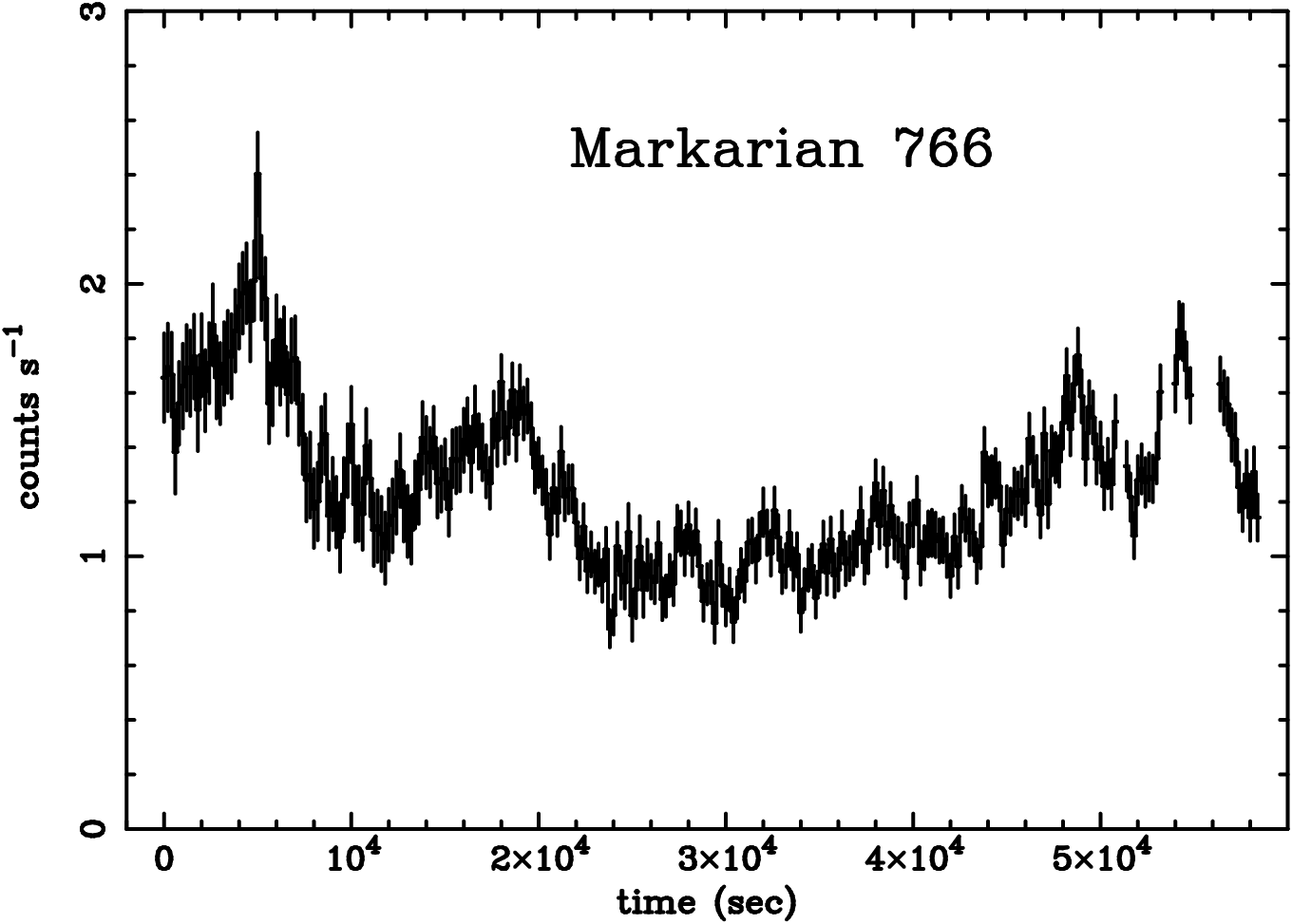}
  \caption{\rgs\ first-order background-subtracted light curve of \mcg\
           (top) and \mrk\ (bottom).  The data are binned to 100 s bins.}
           \label{f1}
\end{figure}

\begin{figure}
  \includegraphics[height=3in,angle=-90]{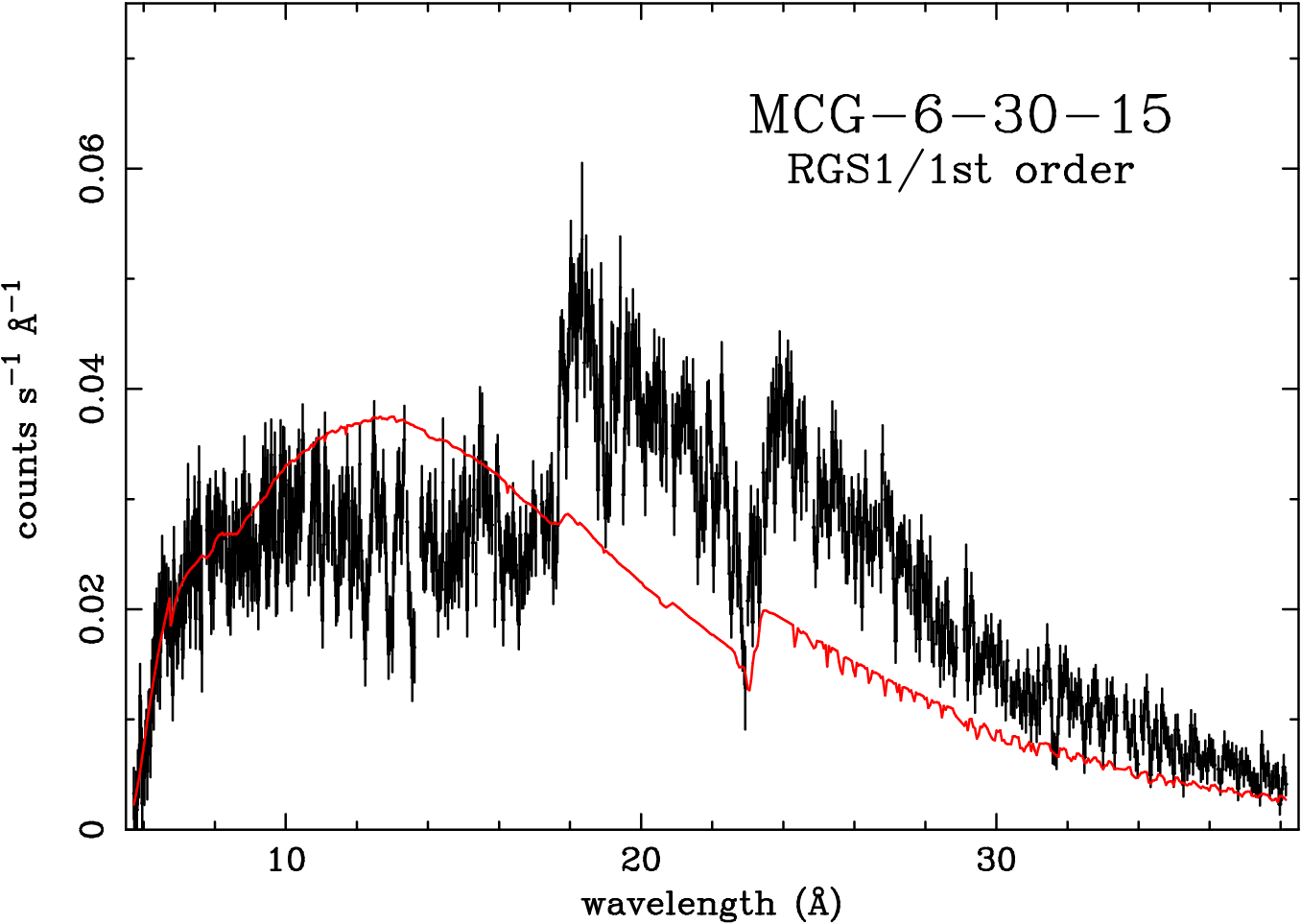}
  \includegraphics[height=3in,angle=-90]{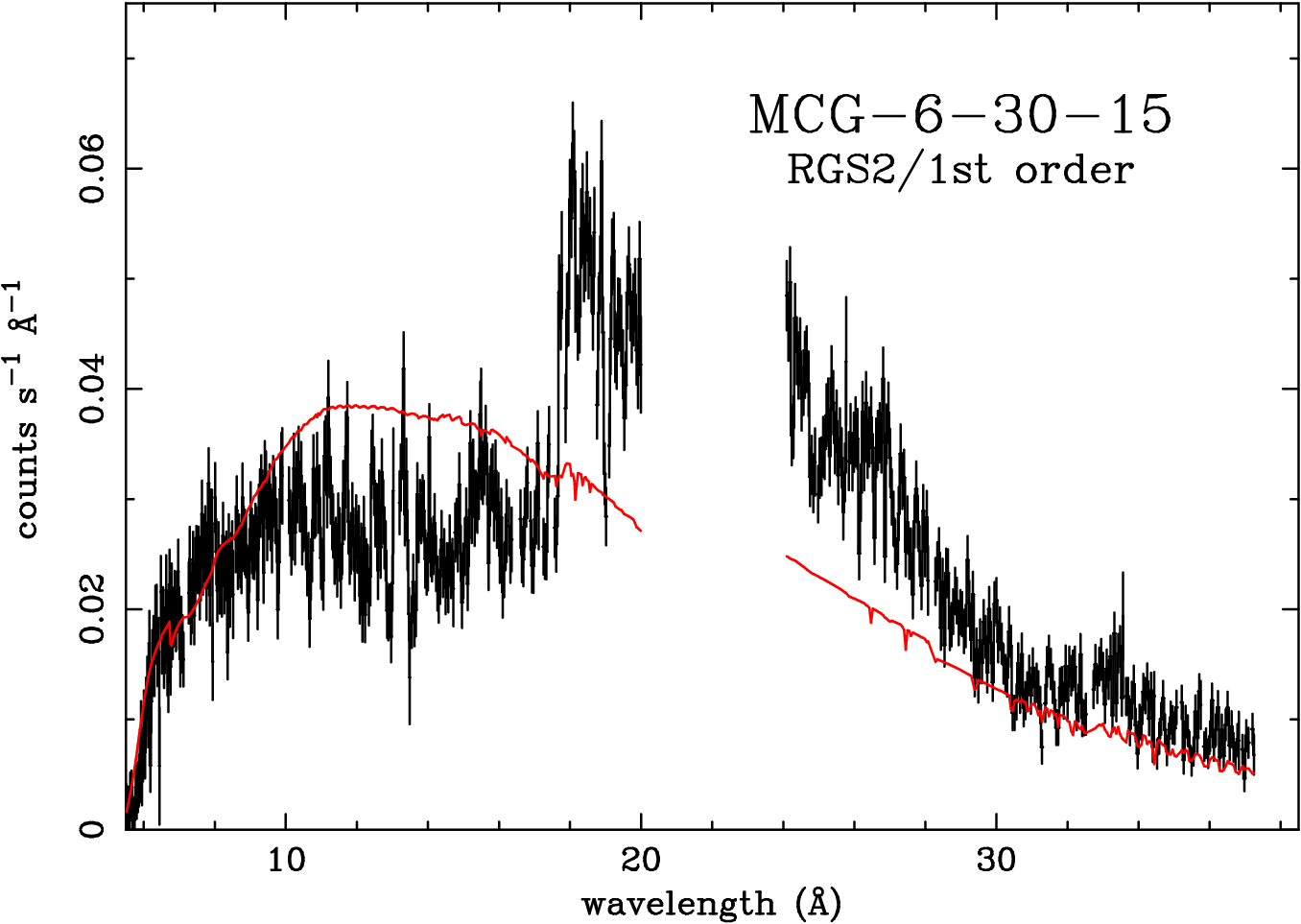}\\
  \includegraphics[height=3in,angle=-90]{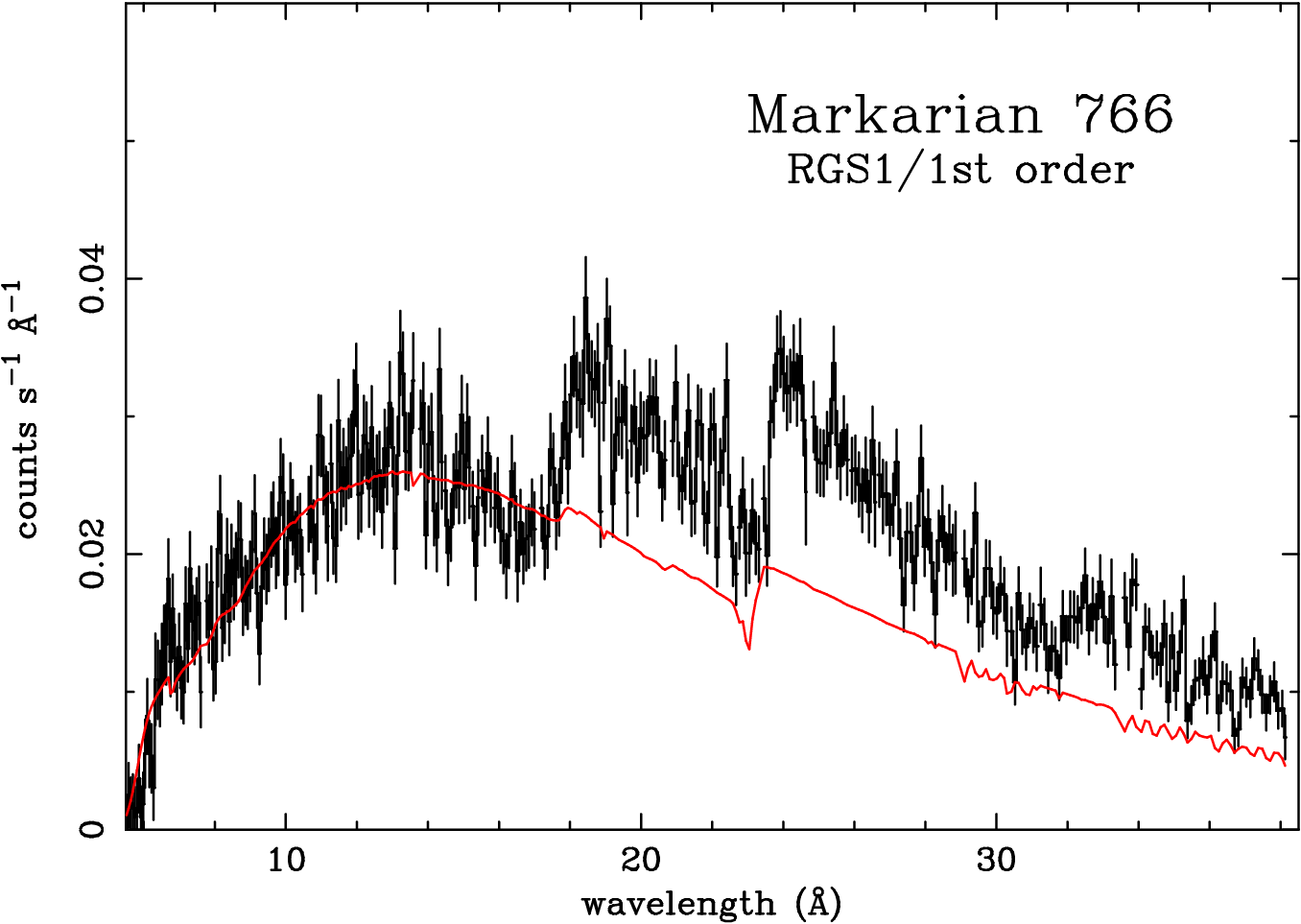}
  \includegraphics[height=3in,angle=-90]{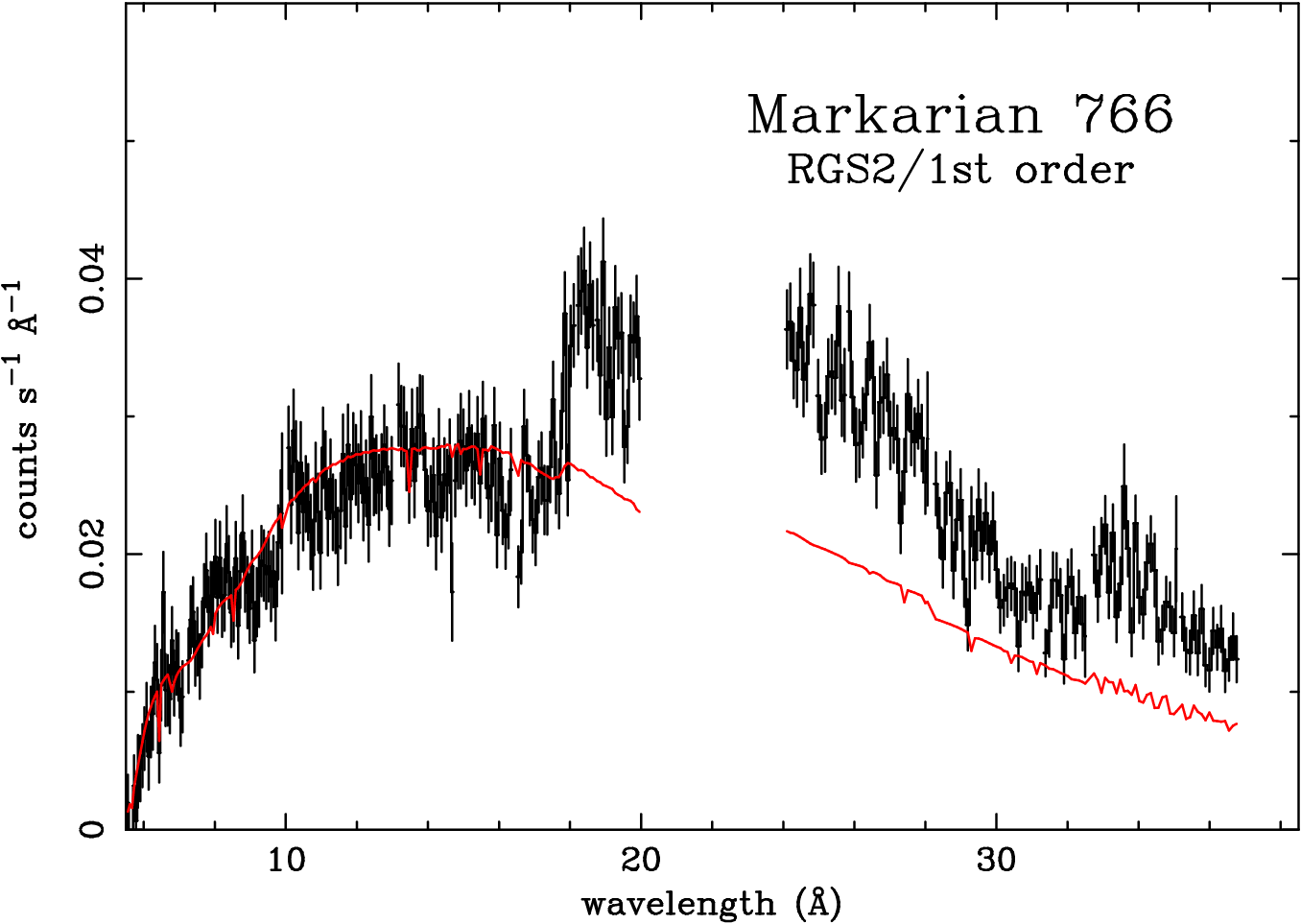}\\

  \caption{\rgs\ spectra of \mcg\ (top) and \mrk\ (bottom) with the \epic\ high
           energy continuum model superimposed (single power law with Galactic
           absorption).  Data from both \rgs1 (left) and \rgs2 (right) are
           shown.  The spectra are rebinned by a factor of 4 relative to their
           respective nominal binsizes (see text) to emphasize the broad
           features.  All figures are plotted in the observers'
           frame.}\label{f2}

\end{figure}

  We first characterize the underlying power law continuum radiation using the
  \epic\ \mos\ and \pn\ data in the $2 - 10 ~\rm{\kev}$ region, excluding the
  region containing the broad iron K line ($5 - 7 ~\rm{\kev}$).  Both sources
  exhibited significant flux variability during the observations, as shown in
  Figure~\ref{f1}.  For the time-averaged spectrum of \mcg, the best fit
  power-law slope is $\Gamma = 1.81 \pm 0.02$ with a normalization of $(8.4
  \pm 0.2) \times 10^{-3} ~\rm{photons} ~\rm{cm}^{-2} ~\rm{s}^{-1}
  ~\rm{\kev}^{-1}$ at 1 \kev.  The integrated flux between $3 - 10 ~\rm{\kev}$
  is $(2.3 \pm 0.1) \times 10^{-11} ~\rm{ergs} ~\rm{cm}^{-2} ~\rm{s}^{-1}$,
  which corresponds to an anomalously low state compared to those studied with
  the \ginga, \asca, \sax, and \xte\ observations of \mcg\
  (\citealt{matsuoka90,nandra90, fabian94,orr97,guainazzi99,lee00}), but
  similar to the ``deep minimum'' state presented by \citet{iwasawa96}.  For
  comparison, we note that the observed spectral index is nearly identical to
  that during the \hetg\ observation of \mcg\ ($\Gamma = 1.84$;
  \citealt{lee01}).  The flux level, however, is $\sim 50$ \% lower during the
  \xmm\ observation.  The time-averaged \pn\ spectrum of \mrk\ shows $\Gamma =
  1.96$ with a normalization of $4.8 \times 10^{-3} ~\rm{photons}
  ~\rm{cm}^{-2} ~\rm{s}^{-1} ~\rm{\kev}^{-1}$ at 1 \kev, which differs
  slightly from the parameters published by \citet{page01} owing to several
  updates in the instrument calibration data.  The \epic\ \pn\ and \mos\ were
  shut off during a large background flare for the first $\sim 18 ~\rm{ksec}$
  of the observation.  Inspection of the background-subtracted \rgs\ data
  during this epoch shows that the source count rate is slightly higher than
  during the remainder of the observation.  The average \rgs\ count rate
  including this epoch is $\sim 10$\% higher than that excluding it.  We,
  therefore, rescale the normalization of the power-law to $5.8 \times 10^{-3}
  ~\rm{photons} ~\rm{cm}^{-2} ~\rm{s}^{-1} ~\rm{\kev}^{-1}$ at 1 \kev.  Shown
  in Figure~\ref{f2} are the derived high energy power-law continua
  superimposed on the \rgs\ data.

  In both sources, the spectra from all three instruments (\rgs, \pn, and
  \mos) show large excess flux in the soft X-ray band above the extrapolation
  of the high energy power-law continuum.  Shortward of $\lambda \sim 17$~\AA\
  ($E \sim 0.7 ~\rm{\kev}$), the spectrum of \mcg\ shows a model excess, which
  is mostly due to the forest of absorption lines by Fe L ions, as discussed
  in detail below.  The spectrum of \mrk\ in the same region shows evidence
  only for very weak absorption.

  The observed excess emission in the soft X-ray band is not an artifact due
  to, for example, spectral softening with increasing flux, as generally
  suggested from previous observations of \mcg\ (\citealt{pounds86,matsuoka90,
  fiore92,lee00}) The data consistently show comparable excess emission
  throughout the entire \xmm\ observation.  Absolute and cross-calibration
  uncertainties are much smaller than the observed excesses.

  \mcg\ is well-known for its broad iron K line, which has been interpreted as
  fluorescent emission from a cold medium, distorted by Doppler and
  gravitational effects in the vicinity of a massive black hole
  \citep{tanaka95,fabian95}.  Recently, however, \citet{wilms01} have shown
  using \epic\ \pn\ data collected simultaneously with the \rgs\ data
  presented here, that the high-energy spectrum of \mcg\ during this epoch is
  consistent with reflection from a highly-ionized relativistic accretion
  disk.  The best-fit model consists of an extremely broad line with a
  rest-frame energy that is consistent with that of H-like iron.  The
  equivalent width $\ew$ of the line is found to be $\sim 300 - 400 ~\rm{\ev}$
  with an extremely steep radial emissivity profile ($\beta \sim 4.7$, where
  the line emissivity $\epsilon \propto r^{-\beta}$).

  We note that the precise determination of the power-law slope is difficult
  for \mcg, since there is essentially no region in the spectrum with a true
  power law.  The soft X-ray region contains various absorption and emission
  features, while the hard X-ray spectrum exhibits a broad iron line.
  \citet{wilms01} have used approximately half of the total exposure time, and
  have shown that $\Gamma = 1.87 - 1.96$ for the various models that provide
  adequate fits to the \epic\ \pn\ spectrum.  Normalizing the flux at $E \sim
  2 ~\rm{\kev}$, where the effects of soft X-ray emission/absorption and the
  contributions from the broad iron line are minimal, the variance in the
  power-law slope produces a flux difference of at most $\sim 10$\% at $E \sim
  0.5 ~\rm{\kev}$.  The observed excess emission is much larger than this value
  and, therefore, our conclusions are not affected qualitatively.

  \mrk\ is known to exhibit rapid, complex spectral and temporal variability
  \citep{molendi93,molendi94,leighly96,page99,matt00,boller01,page01}.  Using
  several observations with \asca\ and \rosat, \citet{leighly96} and
  \citet{page99} have demonstrated that variability is observed mostly in the
  $E = 0.5 - 2 ~\rm{\kev}$ ($\lambda = 5 - 25$~\AA) region, where the spectrum
  is dominated by the power-law component.  The much softer component observed
  in the \rosat\ data, on the other hand, is relatively steady, and it has
  been suggested that it originates from the accretion disk.  \citet{page01}
  have also identified a broad emission line from highly ionized iron near
  $6.5 ~\rm{\kev}$ in the \epic\ data of \mrk.  In contrast to that of \mcg, we
  believe that the power-law slope derived from the \epic\ data of \mrk\ is
  relatively robust, since the observed absorption features are much weaker in
  this source.

  Throughout this paper, we assume that the power-law continuum parameters
  derived from the $2 - 10 ~\rm{\kev}$, excluding the broad iron line region
  ($5 - 7 ~\rm{\kev}$) represent the true high energy continuum shape in these
  sources.  This may not necessarily be an adequate assumption, since we do
  not know a priori whether, for example, the opacity in the warm absorber is
  negligible near $E \sim 2 ~\rm{\kev}$ or whether the broad spectral feature
  in the iron K region is really an emission feature.  Due to the relatively
  complex nature of this problem, we do not explore other possible continuum
  shapes in this paper.  We are, however, investigating these possibilities,
  to be discussed in a future article \citep{ali03}.

\section{Explicit Test of the Lee et al.\ (2001) Model}
  The spectral model for \mcg\ proposed by \citet{lee01} to fit the \chandra\
  \hetg\ data consists of a power-law ($\Gamma = 1.84$) continuum, plus a soft
  component represented by either a faint blackbody with $kT \sim 0.13
  ~\rm{\kev}$ or an additional steep power-law with $\Gamma \sim 2.5$.  The
  model also includes absorption by neutral atoms in dust grains and by highly
  ionized atoms in a warm external medium.  In this section, we investigate
  whether this dusty warm absorber model can also provide an adequate
  representation of the higher statistical quality \rgs\ data, which cover
  also the long wavelength region ($25 \la \lambda \la 38$\AA) outside the
  \hetg\ bandpass.

  The ionized absorption components are characterized as follows.  For each
  discrete transition, we calculate absorption line profiles using oscillator
  strengths and radiative decay rates from \citet{verner96b} for the K-shell
  ions.  The photoelectric opacity is calculated self-consistently as well,
  using cross sections from \citet{verner96a}.  For the oscillator strengths,
  radiative and autoionization rates, and photoelectric cross sections of
  L-shell ions of oxygen and iron, we use our own atomic data, calculated with
  the Hebrew University/Lawerence Livermore Atomic Code (\hullac;
  \citealt{barshalom01}).  Decay rates and oscillator strengths calculated
  with \hullac\ may be inaccurate by as much as $\sim 20$\%, and the
  wavelengths are accurate to within $\sim 50$ m\AA.  For the strong Fe L
  resonance lines, we use laboratory transition wavelength measurements of
  \citet{brown02}.  We also use values published by \citet{pradhan00} for the
  strong \ion{O}{6} inner-shell absorption line properties.  Each ion is
  treated as a single component in the spectral fit, and is characterized by
  three parameters: (1) the ion column density $N_i$, (2) the turbulent
  velocity width or $b$-parameter defined as $b = \sqrt{2}\sigma$, where
  $\sigma$ is the Gaussian width, and (3) the Doppler velocity shift.  We note
  that the lines in the data appear unresolved with both the \rgs\ and the
  \hetg\ and, therefore, the $b$-parameter cannot be measured directly.  The
  equivalent width ratios for lines from a given ion are, however, sensitive
  to the intrinsic line width, so we fit all lines from a given ion
  simultaneously, even in cases where some of the lines are saturated.

  This method, in most cases, allows for rather tight constraints to be
  derived individually on both $N_i$ and the $b$-parameter.  At column
  densities of $N_i \la 10^{16} ~\rm{cm}^{-2}$, the lines are at most mildly
  saturated and so the equivalent widths increase linearly with column density
  and do not depend on the $b$-parameter.  In this limit, the column density
  can be determined with a relative uncertainty that is smaller than the
  uncertainties in the measured $\ew$, but the $b$-parameter is unknown.  For
  $N_i \ga 10^{17} ~\rm{cm}^{-2}$, the change in the $\ew$ is due to
  absorption in both the Doppler and damping wings\footnote{Note that since
  the transition probabilities of X-ray lines are much larger than those of
  typical optical and UV lines by orders of magnitude, absorption in the
  damping wings become important at a relatively lower column density.  As a
  consequence, strong X-ray resonance lines have only a very narrow range in
  column density where the curve of growth is logarithmic.} and, therefore,
  the $\ew$ ratio depend sensitively on both the column density and the
  $b$-parameter.

  This spectral fitting method is purely empirical, and unlike global models
  that are typically characterized by an ionization parameter, equivalent
  hydrogen column density, velocity field, etc., we make no assumptions
  regarding the ionization and thermal balance of the absorbing gas.  Hence,
  the number of parameters is large (each ion column density is a free
  parameter).  On the other hand, since the assumptions are minimal, the
  fitting procedure is computationally efficient and convergence is achieved
  typically on the order of minutes.  In addition, for components that are
  unresolved in velocity space, each ion is characterized by a single column
  densitiy, which may or may not be an adequate description of the true
  physical situation as there may be numerous spatially distinct regions in
  the absorber with the same outflowing velocity but different turbulent
  velocities.  The data, however, are not sensitive to the various possible
  physical scenarios.

  With the exception of the \ion{O}{7} He-like triplet region, we ignore the
  effect of re-filling of absorption lines from resonance line photons that
  scatter into our line of sight.  This is an excellent approximation for most
  transitions with upper levels that lie above the lowest excited state (e.g.,
  He$\beta$, He$\gamma$, etc.).  For the lowest transitions (i.e.,
  He$\alpha$), however, this approximation may not be entirely valid, as
  discussed in the following section.  We check for self-consistencies
  whenever possible.  We finally note that possible line emission from
  collisionally ionized plasma is likely to be completely negligible in
  filling in the observed absorption lines, as it requires an unusually large
  emission measure of $\ga 10^{64} ~\rm{cm}^{-3}$.  This is $\sim 2$ orders
  magnitude more than observed in even the most intense starburst galaxies
  (e.g., NGC~253, \citealt{pietsch01}; M82, \citealt{read02}) .

\begin{figure}
  \includegraphics[height=6in,angle=-90]{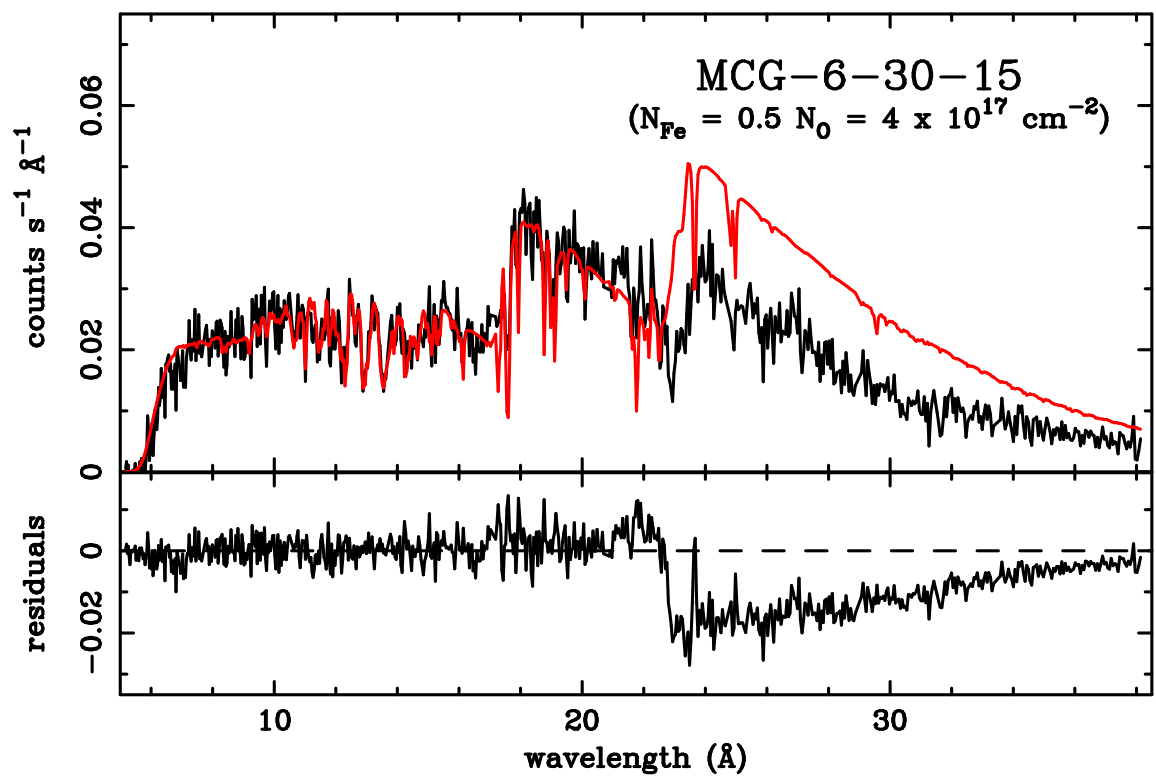}
  \caption{The model proposed by \citet{lee01} superimposed on the \rgs\
           data.  The blackbody flux is normalized to the data
           near $\lambda \sim 18$~\AA.  Note that the general properties of
           the spectrum covered by the \hetg\ bandpass (below $\lambda
           \sim 23$~\AA) are well-reproduced, although the details of the
           fit, particularly the \ion{O}{7} absorption line equivalent
           widths, are not.}
           \label{f3}
\end{figure}

  We first test a model in which the power-law index and normalization are
  fixed to those derived from the \epic\ data, and the blackbody temperature
  to $kT = 0.13 ~\rm{\kev}$ \citep{lee01}.  Since the normalization of the
  blackbody component is not quoted in \citet{lee01}, we scale the flux to the
  measured spectrum near $\lambda = 18$~\AA.  The flux in the blackbody
  component corresponds to a circular surface area with a radius of $R_{\rm
  bb} \sim 2 \times 10^{10} ~\rm{cm}$.  Ionic column densities and the
  $b-$parameter values are allowed to vary in the fit.  The resulting fit is
  shown in Figure~\ref{f3}.  As is evident from the figure, the model fits the
  overall shape of the spectrum fairly well below $\lambda \sim 23$~\AA, i.e.,
  in the region of overlap between the \rgs\ and \chandra\ \hetg\ bandpasses.
  This model, however, grossly overestimates the flux above $\lambda \sim
  25$~\AA, as well as the equivalent widths of the oxygen absorption lines.
  The mismatch between the model and the data is largest just above the
  neutral oxygen edge and gradually improves towards longer wavelengths (see
  Figure~\ref{f3}).  The energy dependence of the residuals cannot be modeled
  as some form of additional absorption, since photoelectric cross sections
  decrease towards shorter wavelengths.  A steeper power-law in place of the
  blackbody component provides an even worse representation of the data, as it
  produces more flux towards longer wavelengths than a blackbody.

  \citet{lee01} interpret the sharp drop near $\lambda \sim 17.5$~\AA\ as a
  combination of \ion{O}{7} ($\lambda = 16.78$~\AA) and neutral iron L-shell
  ($\lambda = 17.55$~\AA) absorption.  Although the general characteristics of
  their predicted feature are indeed close to what we observe, this
  interpretation cannot explain the detailed shape of the spectrum.  The
  corresponding \ion{O}{7} absorption line equivalent widths are
  overpredicted, as discussed in detail in \S4.1.1.  If we assume that the
  underlying continuum is smooth across this feature as in \citet{lee01}, the
  model flux between the \ion{O}{7} He$\delta$ ($\lambda = 17.42$~\AA) and
  He$\epsilon$ ($\lambda = 17.22$~\AA) lines lies above the data, owing to the
  lack of \ion{O}{7} and neutral iron opacity in this narrow wavelength
  region.  Additional velocity broadening improves the fit somewhat, however,
  at the expense of overpredicting the He$\alpha$, $\beta$, and $\gamma$
  absorption $\ew$.  We find that absorption by neutral iron is required by
  the data, but with a much lower column density than posited by \citet{lee01}
  (see \S4.2).  The data, however, are much better fit if the flux of the
  underlying continuum emission increases gradually from $\lambda \sim
  17.0$~\AA\ to $\lambda \sim 17.6$~\AA.

\begin{figure}
  \includegraphics[height=4in,angle=-90]{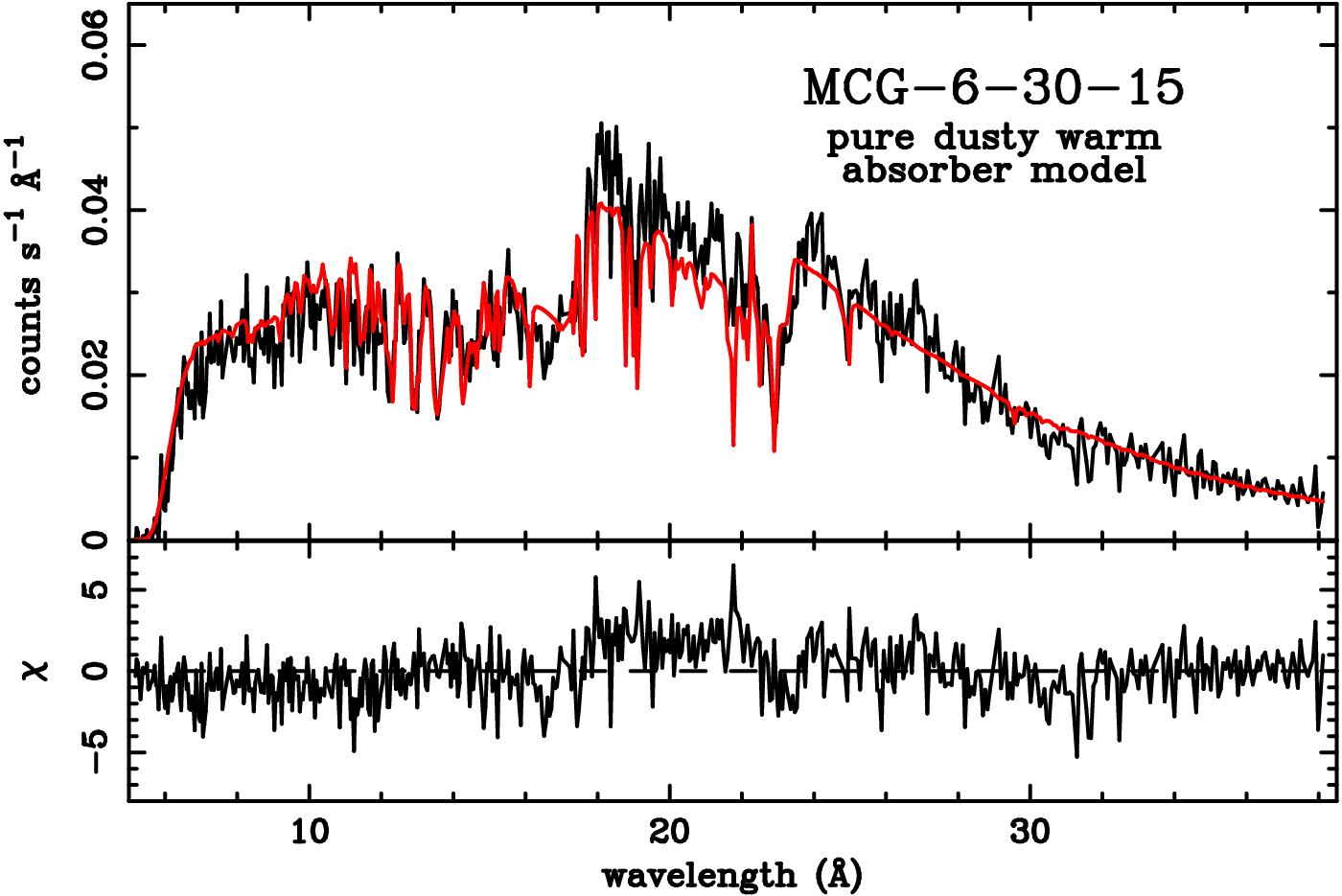}\\
  \includegraphics[height=4in,angle=-90]{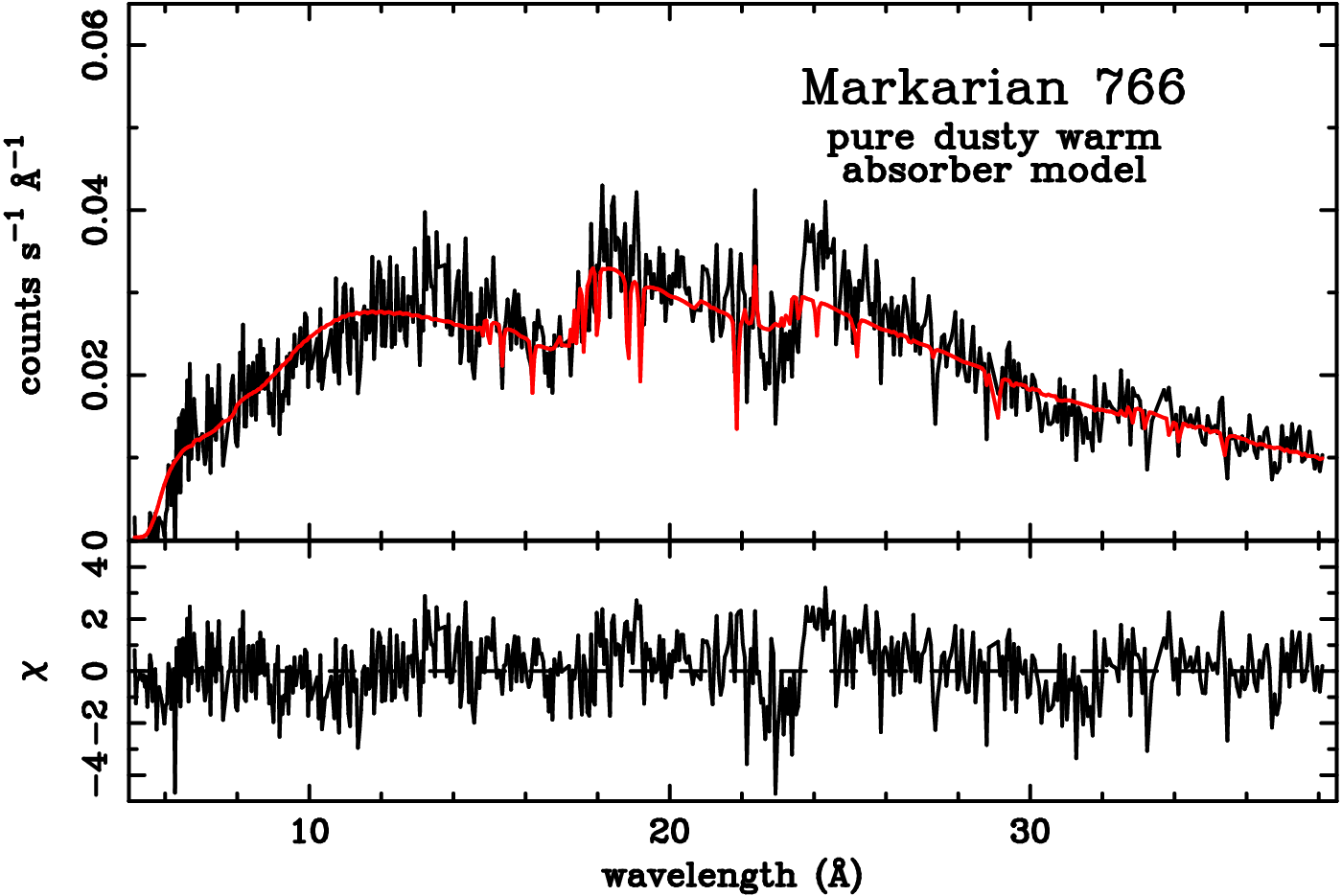}
  \caption{Best-fit dusty warm absorber model for \mcg\ (top) and \mrk\
           (bottom).  Broad residuals are seen in both spectra, and
           cannot be explained by any absorption features.}
           \label{f4}
\end{figure}

  When we allow the blackbody temperature and normalization to be free, the
  fit improves dramatically.  The details of the spectrum, however, cannot be
  well-reproduced, and the model still leaves broad features in the residuals
  that are significantly above the noise level of our data, as shown in
  Figure~\ref{f4}.  In addition, the absorption line equivalent widths,
  particularly of \ion{O}{7}, are overpredicted by the model.  Therefore, a
  pure dusty warm absorber model with a power law plus blackbody emission as
  the intrinsic continuum cannot explain the overall spectrum, even if we
  allow for arbitrarily high ion column densities that produce observable
  photoelectric edges and, in addition, overestimate the absorption line
  equivalent widths.  Again, a broken powerlaw that steepens in the softer
  band ($\lambda \ga 15$~\AA) instead of a blackbody component provides a
  worse representation of the data.

\subsection{Detailed Measurements of the Ionized Absorption Parameters}
  We now turn to describe our own parametrization of the ionized absorber in
  \mcg\ and \mrk, following the procedures described in \S4.  

\subsubsection{\mcg}
\begin{figure}
  \includegraphics[height=6in,angle=-90]{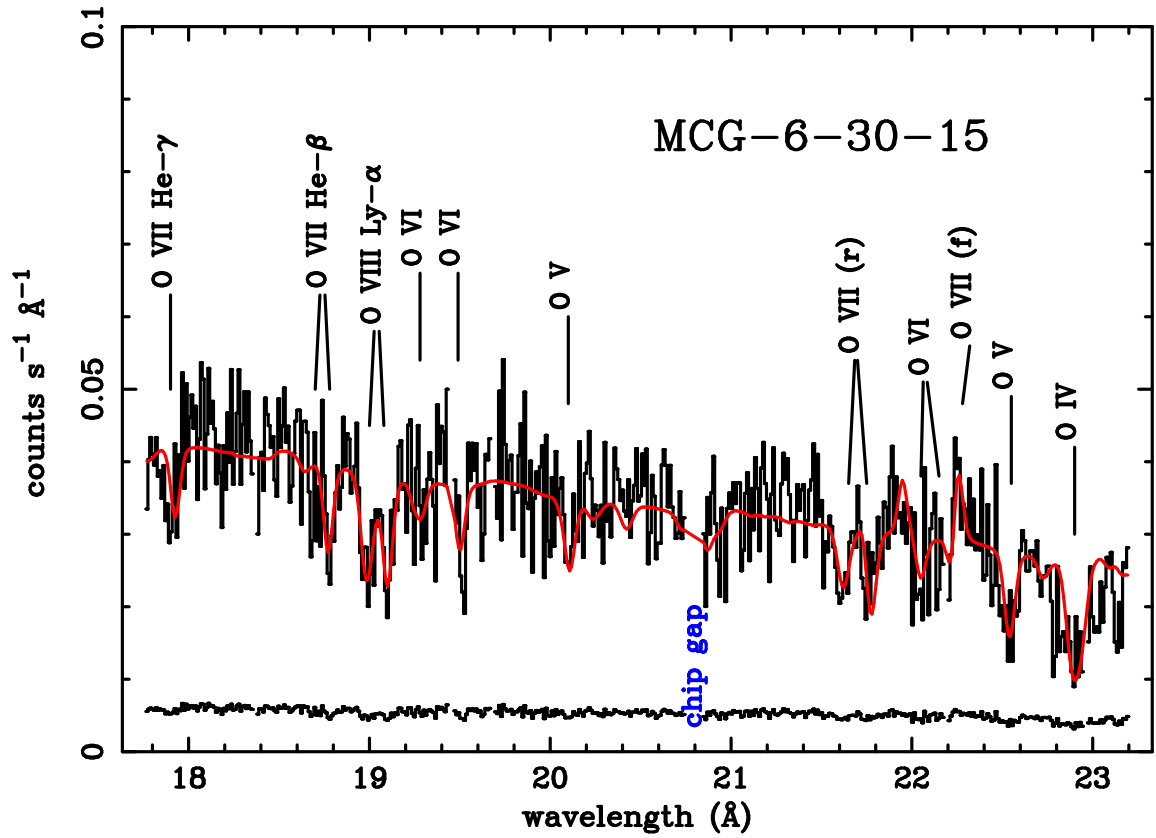}
  \caption{\mcg\ in the oxygen line region.  Absorption and emission lines
           from \ion{O}{8} -- \ion{O}{4} are clearly detected in the
           spectrum.  The He$\gamma$ line near $\lambda = 17.8$~\AA\ is
           contaminated by the instrumental fluorine edge.} \label{f5}
\end{figure}

  The \ion{O}{8} Ly$\alpha$ ($\lambda = 18.969$~\AA) and \ion{O}{7} He$\alpha$
  ($\lambda = 21.602$~\AA) absorption line profiles show at least two distinct
  kinematic components, both of which are blueshifted with outflowing
  velocities of $v_1 = -150 \pm 130 ~\rm{km~s}^{-1}$ and $v_2 = -1900 \pm 140
  ~\rm{km~s}^{-1}$ (hereafter, components 1 and 2, respectively), where the
  errors consist of statistical as well as systematic uncertainties in the
  wavelength calibration.  Multiple velocity components have been observed in
  the X-ray band also in \iras\ \citep{sako01} and in {\small NGC}~4051
  \citep{collinge01}.  We first derive the column densities and turbulent
  velocities from the ionized oxygen lines in the $17.5 - 23.0$~\AA\ band.
  This region of the spectrum contains the resonance transitions of \ion{O}{8}
  Ly$\alpha$, He$\alpha$, $\beta$, and $\gamma$ lines of \ion{O}{7}, as well
  as all of the inner-shell resonances of \ion{O}{6} - \ion{O}{4}.  For the
  present purpose, we do not use the region of the spectrum shortward of $\sim
  17.5$~\AA, since the modeling of the underlying continuum across this
  wavelength range is not a straightforward task, and the only detectable
  absorption feature, in any case, is the \ion{O}{8} Ly$\beta$ line.
  Absorption lines from all of the ions listed above are present in the data
  as shown in Figure~\ref{f5}, and they collectively provide constraints on
  the turbulent velocity width and the individual ion column densites.  The
  bulk velocity shifts and turbulent velocity widths of each oxygen ion, with
  the exception of \ion{O}{4}, are tied, and are allowed to vary in the fit.
  The \ion{O}{4} absorption line complex cannot be fit by the velocity shift
  and width derived from the \ion{O}{8} - \ion{O}{5} lines, which we believe
  is mainly due to the inaccuracy of the present wavelength calculations of
  the order $\sim 50$~m\AA.  The spectrum also shows a weak, but statistically
  significant, forbidden emission line of \ion{O}{7} at $\lambda = 22.10$~\AA.
  We use this line to normalize the amount of \ion{O}{7} recombination
  emission, which can partially fill the resonance absorption line, as well as
  the higher series absorption lines.  We empirically model the underlying
  continuum locally as a single power law with Galactic absorption.  The
  best-fit value for the slope is $\Gamma = 1.6$ with a normalization of $1.6
  \times 10^{-2} ~\rm{photons} ~\rm{cm}^{-2} ~\rm{s}^{-1} ~\rm{\kev}^{-1}$ at
  1 \kev, which is slightly flatter and much brighter than that of the high
  energy region determined from the \epic\ data.  This implies the presence of
  an additional spectral component, which is flatter than the primary
  power-law.  We emphasize that the continuum parameters derived above are
  purely empirical, and are used only for the sake of measuring the absorption
  parameters.

\begin{figure}
  \includegraphics[height=4in,angle=-90]{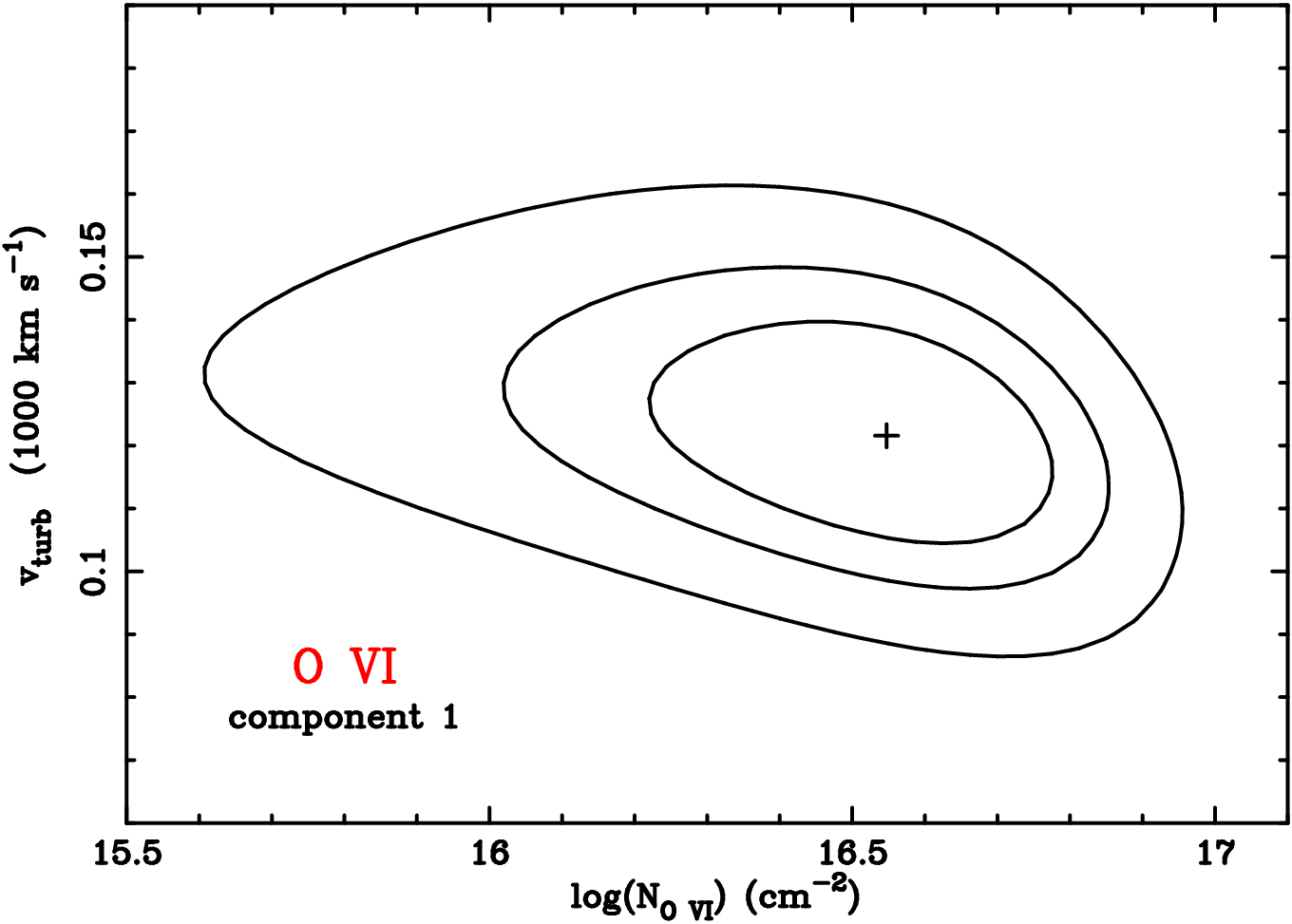}\\
  \includegraphics[height=4in,angle=-90]{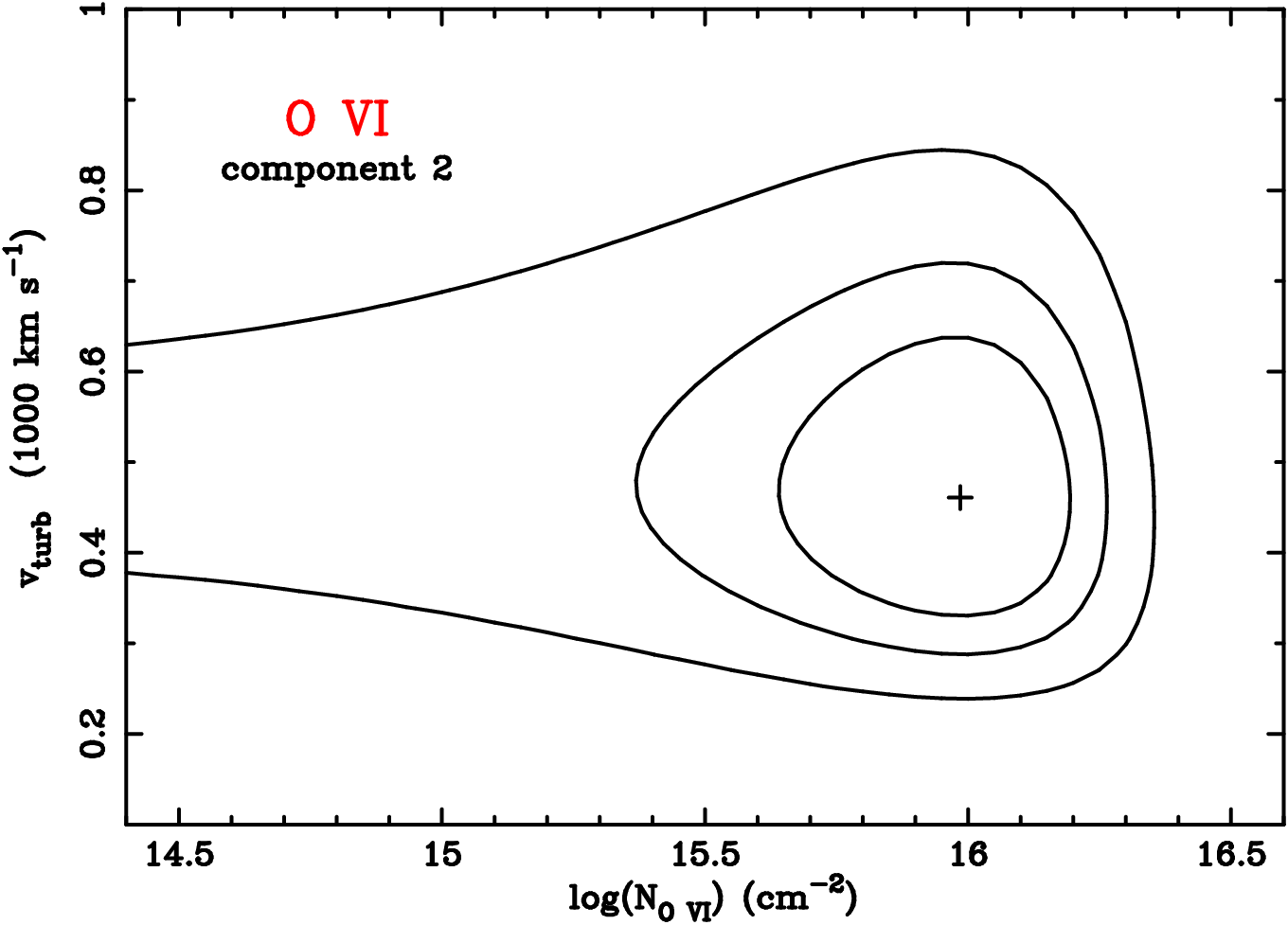}
  \caption{Confidence countour regions in turbulent velocity width and
           column density for \ion{O}{6} for the low-velocity
           component (top) and the high-velocity component (bottom) in \mcg.}
           \label{f7}
\end{figure}

\begin{figure}
  \includegraphics[height=4in,angle=-90]{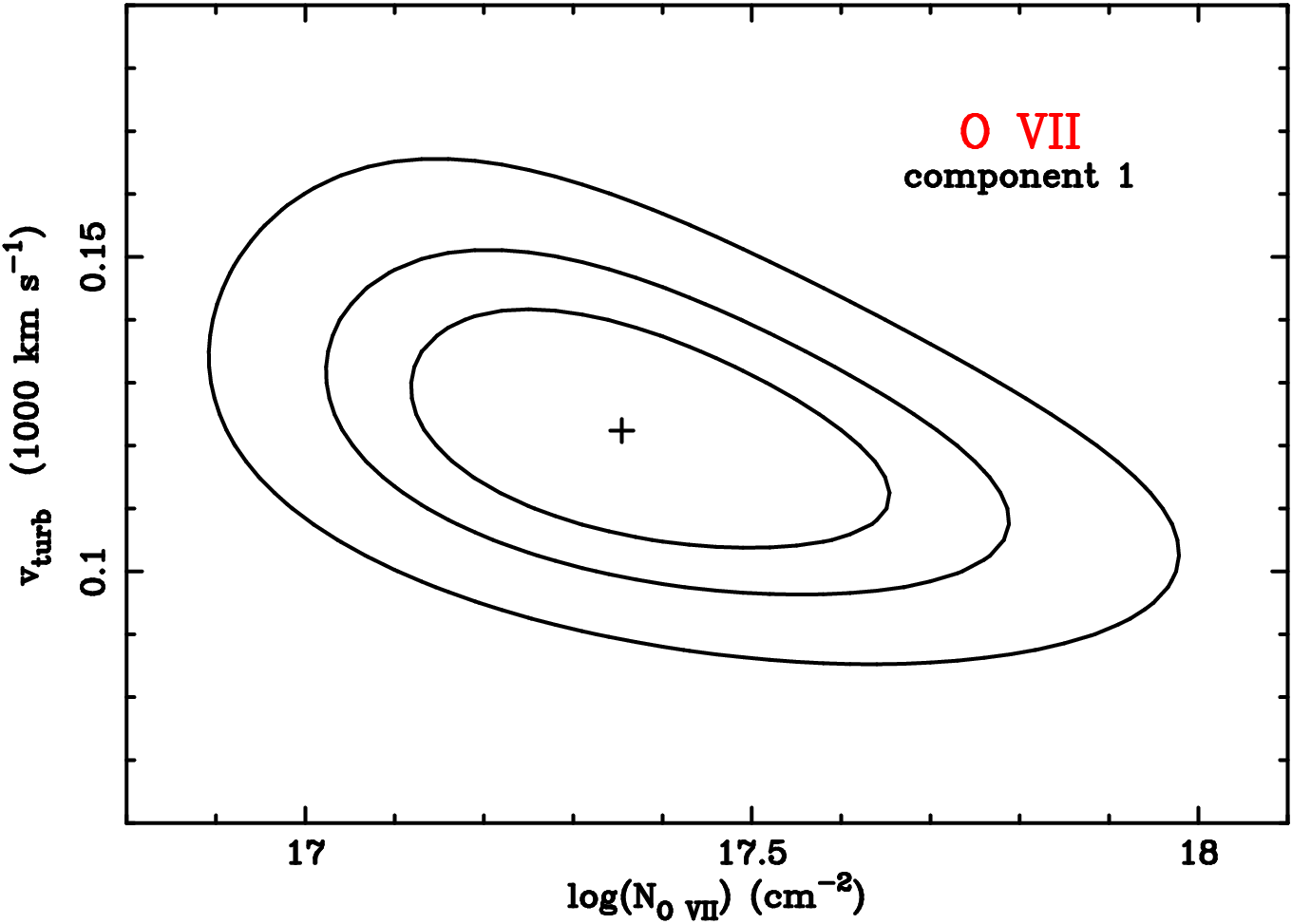}\\
  \includegraphics[height=4in,angle=-90]{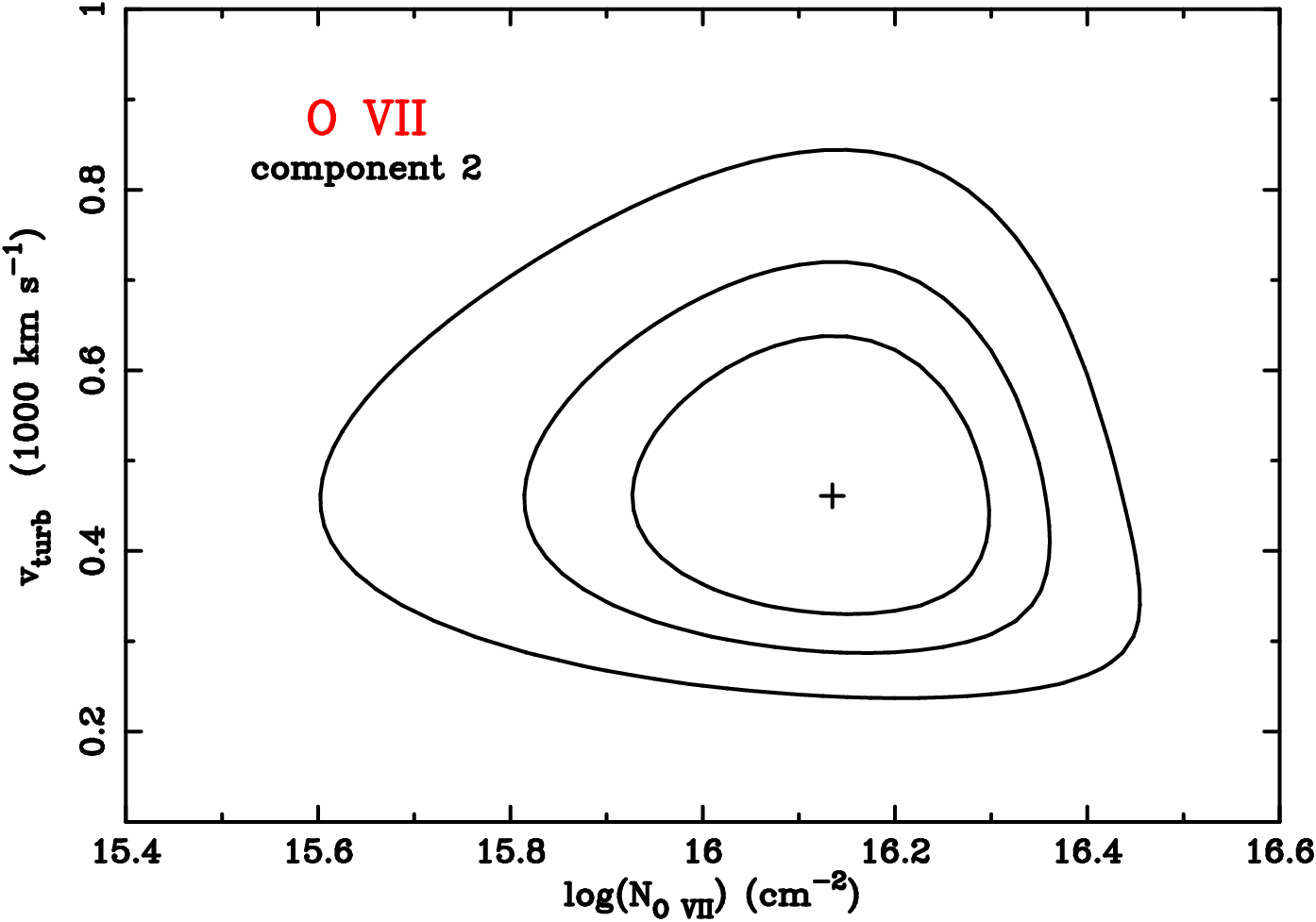}
  \caption{Same as in Figure~\ref{f7} for \ion{O}{7} for the low-velocity
           component (top) and the high-velocity component (bottom).}
           \label{f8}
\end{figure}

\begin{figure}
  \includegraphics[height=4in,angle=-90]{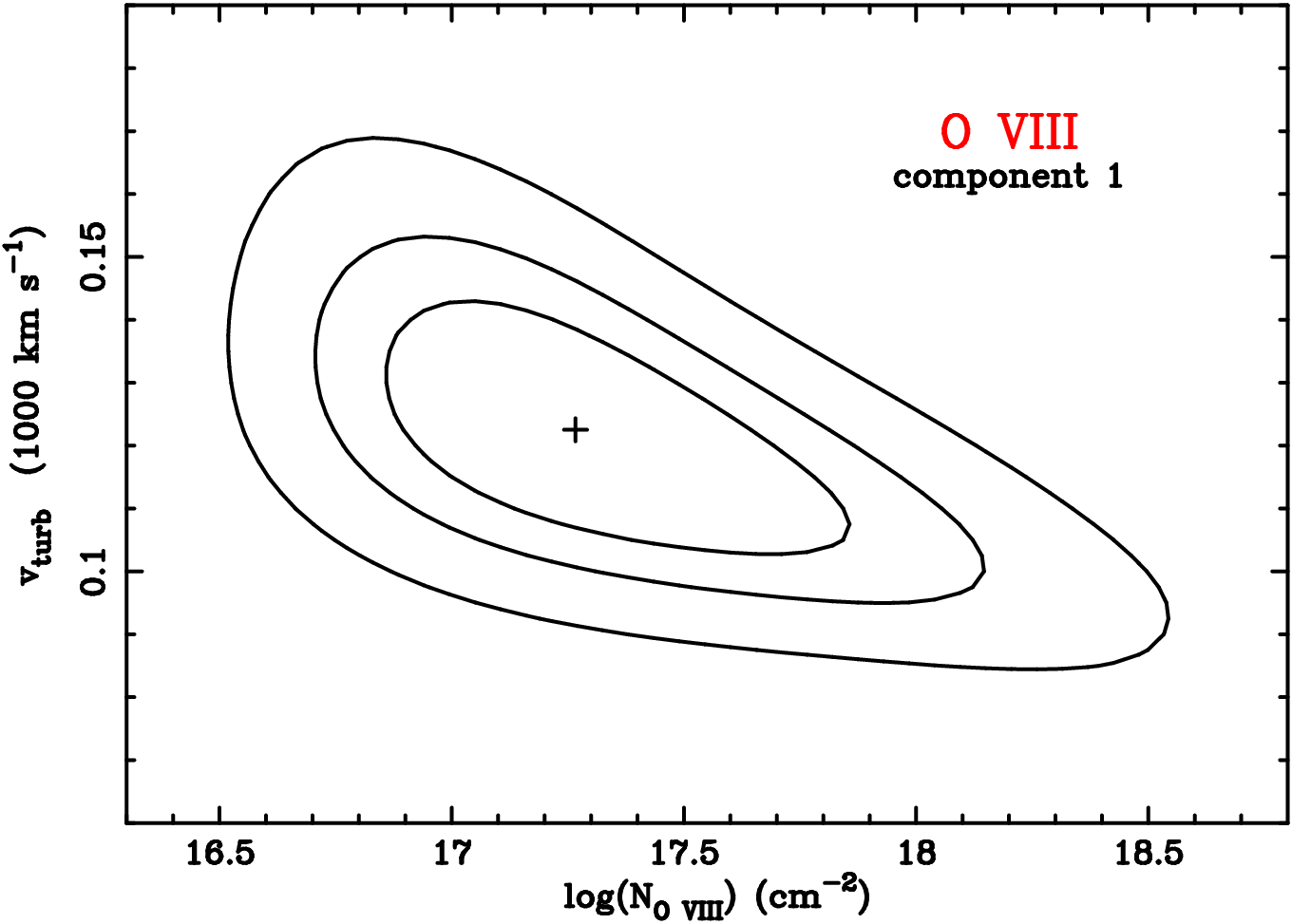}\\
  \includegraphics[height=4in,angle=-90]{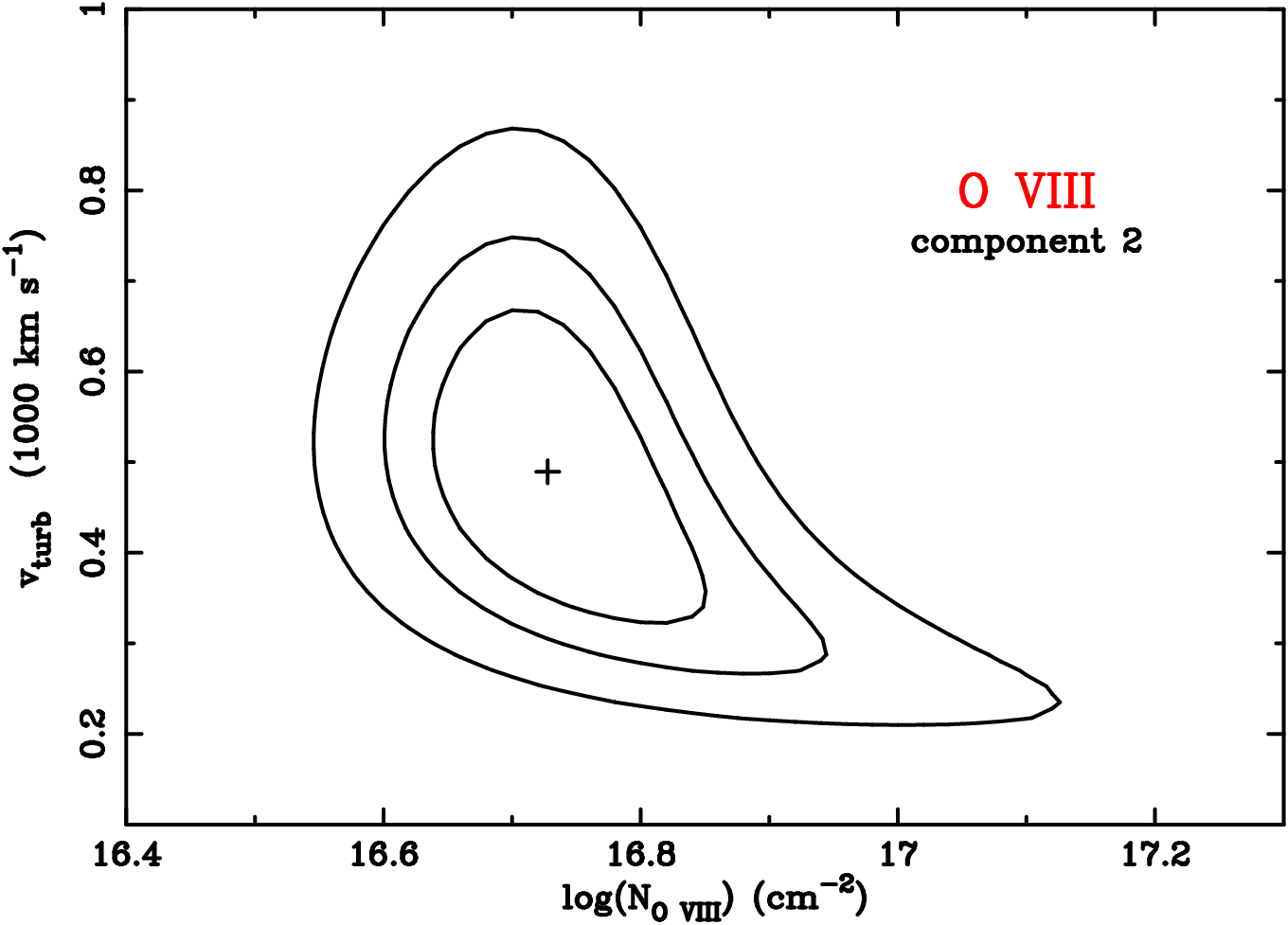}
  \caption{Same as in Figure~\ref{f7} for \ion{O}{8} for the low-velocity
           component (top) and the high-velocity component (bottom).}
           \label{f9}
\end{figure}

  The measured oxygen ion column densities for the two kinematic components
  are listed in Table~\ref{tbl2}.  The best-fit turbulent velocity widths are
  found to be $b_1 = 130 ~\rm{km~s}^{-1}$ and $b_2 = 460 ~\rm{km~s}^{-1}$.
  Also shown in Figures~\ref{f7}, \ref{f8}, and \ref{f9} are the 68, 90, and
  99\% two-parameter confidence regions for the $b$-parameter and the ion
  column densities of \ion{O}{6} -- \ion{O}{8}.  The \ion{O}{7} column
  densities are particularly important for comparison to the \citet{lee01}
  model.  For the two components, we obtain $N_{1}$(\ion{O}{7}) $= 2.2 \times
  10^{17} ~\rm{cm}^{-2}$ and $N_{2}$(\ion{O}{7}) $= 1.3 \times 10^{16}
  ~\rm{cm}^{-2}$, both of which are lower by at least an order of magnitude
  than what is required to produce a photoelectric absorption edge with $\tau
  \sim 0.7$ at threshold; a value claimed by \citet{lee01} to explain the
  spectral jump near $\lambda \sim 17$~\AA.  The 99\% upper limit of component
  1 is still a factor of $\sim 3$ lower.  The data, therefore, invalidate any
  model that abscribes the turnover of the spectrum at $\lambda \sim
  17.5$~\AA\ to \ion{O}{7} photoelectric opacity.

\begin{figure}
  \includegraphics[height=6in,angle=-90]{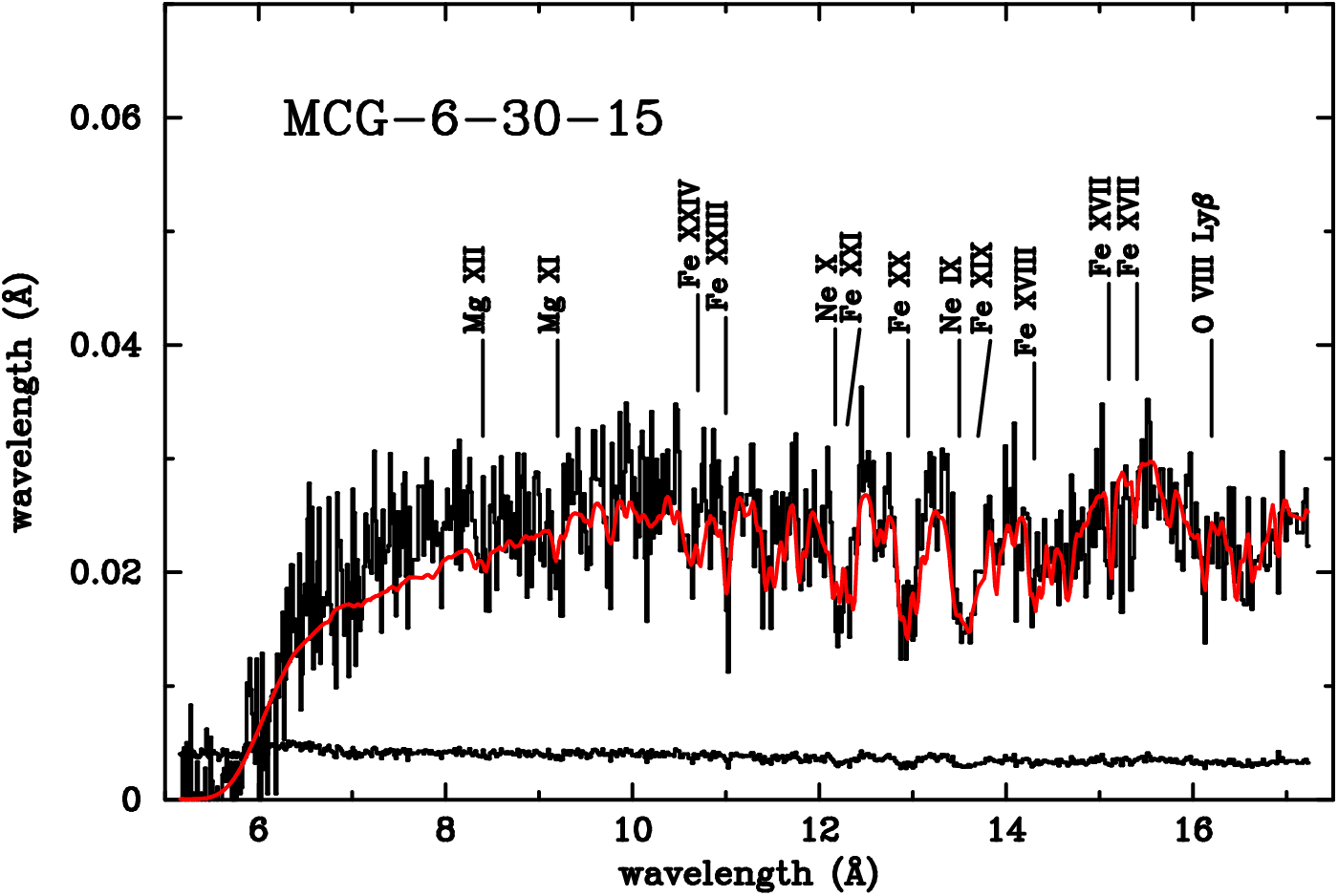}
  \caption{The short wavelength region of the \mcg\ data with the best-fit
           model superimposed.  The broad absorption feature in the
           $16 - 17$~\AA\ region is due to the presence of a weak iron \uta\
           (\citealt{chenais00,sako01,behar01}).  Note that this feature
           cannot be explained by an \ion{O}{7} photoelectric edge, since
           (1) the strengths of the absorption lines do not predict an
           observable edge and (2) the shape is inconsistent with that of an
           edge.  As in Figure~\ref{f5}, only data from \rgs1 are shown and
           the lower curve represents 1$\sigma$ Poisson fluctuations.}
           \label{f10}
\end{figure}

  Using the velocity shifts and widths derived from the oxygen absorption
  lines as described above, we measure the column densities of ions with
  discrete transitions in the $5 - 17$~\AA\ region.  Absorption lines from
  \ion{Ne}{9}, \ion{Ne}{10}, \ion{Mg}{11}, \ion{Mg}{12}, and \ion{Fe}{17} --
  \ion{Fe}{24} are clearly detected in the spectrum.  The resulting fit is
  shown in Figure~\ref{f10}, and the measured column densities for these ions
  are also listed in Table~\ref{tbl2}.  A typical one-parameter 90\%
  confidence range for the column density is $\sim 0.3$ dex.  There is a broad
  absorption trough in the $16 - 17$~\AA\ region.  This is probably due to the
  presence of a weak ($N \la 10^{16} ~\rm{cm}^{-2}$) unresolved transition
  array (\uta) of iron \citep{chenais00, sako01, behar01}, which is also
  included in the model.  Note that many of the absorption lines are blended,
  but the procedure of self-consistently fitting the entire ion absorption
  spectrum for each charge state allows for a robust determination of the
  individual ion column densities.

  We note that the velocity width of component 2, which was derived from the
  observed equivalent width ratios of the oxygen absorption lines, may
  possibly be due to the presence of additional discrete kinematic components
  that are unresolved with the \rgs.  This is possible, since the derived ion
  column densities of \ion{O}{8} -- \ion{O}{6} in component 2 are such that
  the strongest resonance lines are still in the optically thin limit, and,
  therefore, the absorption equivalent widths for all of the lines increase
  linearly with line optical depth.  On the other hand, component 1 most
  likely consists of a single velocity component, because the strongest lines
  are saturated, and the measured equivalent width ratios provide unique
  column densities for each of the ions.

  From the relative distributions of the ion column densities listed in
  Table~\ref{tbl2}, the range in ionization parameter ($\xi = L_X/(nr^2)$,
  where $L_X$ is X-ray luminosity, $n$ is the proton number density, and $r$
  is the distance from the continuum source) of component 1 is $0.5 \la \log
  \xi \la 2$, based on calculations with the photoionization code XSTAR
  \citep{kallman95}.  Although the derived range is only approximate, it is
  clear that a single-zone model cannot explain, for example, the observed
  distribution of the oxygen and iron column densities.  As mentioned briefly
  above, both velocity components could, in principle, consist of more than
  one spatially distinct region, which cannot be resolved with the present
  data.  Assuming that the iron abundance in \mcg\ is roughly twice solar
  \citep{lee99}, the estimated equivalent hydrogen column density of component
  1 is $N_{\rm H}(1) \sim 2 \times 10^{21} ~\rm{cm}^{-2}$.  Similarly, for
  component 2, the approximate range in ionization parameter is $2 \la \log
  \xi \la 3$ with a column density of $N_{\rm H}(2) \sim 2 \times 10^{21}
  ~\rm{cm}^{-2}$.  Again, a single-zone model cannot explain the observed
  column density distribution.

  The presence of a broad range in ionization parameter has also been inferred
  from X-ray observations of the Seyfert~2 galaxy {\small NGC}~1068 (see,
  e.g., \citealt{ali02}).  In addition, spatially-resolved spectroscopy with
  the \chandra\ grating spectrometers of the same source shows that the
  ionization structure of the extended photoionized medium is not radially
  stratified, but, instead, consists of a broad range in ionization parameter
  at various radii from the central continuum source, suggesting that the
  medium is highly clumped with density inhomogeneities that span at least a
  few orders of magnitude \citep{brinkman02,ogle03}.  Unfortunately in
  Seyfert~1 galaxies, the warm absorber is spatially unresolved and, hence,
  its location relative to the central black hole is entirely unknown.  It is
  highly likely, however, that the density and ionization structures are
  similar to those seen in Seyfert~2 galaxies.

  There are important similarities, as well as differences, in the details of
  the absorption measurements between those quoted in \citet{lee01} using the
  \chandra\ \hetg\ and our measurements using the \rgs\ data.  The measured
  equivalent width, $\ew$, of the \ion{O}{7} He$\alpha$ resonance absorption
  line in the \chandra\ data is $17 \pm 9$ m\AA, and is completely compatible
  with the \rgs\ data, which shows $\ew = 18 \pm 3$ m\AA.  The measured
  \ion{O}{7} He$\alpha$ forbidden line $\ew$ in the \rgs\ data accounting for
  partial absorption by \ion{O}{6}, however, is $29 \pm 5$ m\AA, which is a
  factor of $\sim 3$ lower than that seen in the \hetg\ spectrum, but probably
  within their statistical uncertainty.  This is, again, due to the low
  signal-to-noise ratio of the \hetg\ spectrum, which contains only a few
  counts per resolution element ($\Delta \lambda \sim 15$ m\AA) in the oxygen
  region.

  There is a limit to the amount of resonance absorption line $\ew$ that can
  be filled in, given the measured $\ew$ of the forbidden emission line.  The
  forbidden line in a photoionized He-like ion is produced solely through
  recombination and subsequent cascades and, therefore, its line intensity
  provides a measure of the He-like ion photoionization rate that occur in the
  plasma.  The resonance line, on the other hand, is affected by discrete
  photoexcitation by the continuum radiation field as well, and the measured
  line flux has contributions from both cascades following recombination and
  radiative decay following photoexcitation.  Resonance line cross sections,
  however, are several orders of magnitude larger than photoionization cross
  sections.  Since the resonance lines saturate at much lower column
  densities, the forbidden to resonance line ratio, therefore, provides a
  sensitive measure of the total ion column density through the emission
  region for a given velocity width.

  If the column density through the medium is high, such that $\tau \sim 0.7$
  at the photoelectric edge as quoted in \citet{lee01}, the number of
  photoionizations dominates over that of photoexcitations in producing the
  resonance line, and the forbidden to resonance ($f/r$) line ratio is close
  to what one expects from pure recombination, as in the calculations of
  \citet{porquet00}.  Recalling that the {\it measured} equivalent width of
  the resonance absorption line of \ion{O}{7} is $18$ m\AA\ and that of the
  forbidden line in emission is $29$ m\AA, and using the fact that the $f$ to
  $r$ ratio in a recombination-dominated plasma is $\sim 3$, the estimated
  {\it total} absorption $\ew$ of the $r$ line, after subtraction of the
  re-emitted resonance line radiation, is $\sim 28$ m\AA.  This corresponds to
  an \ion{O}{7} ion column density of $N_{\rm OVII} \sim 6 \times 10^{16}
  ~\rm{cm}^{-2}$.  That is, again, more than an order of magnitude lower than
  $\sim 3 \times 10^{18} ~\rm{cm}^{-2}$ ($\tau \sim 0.7$), and is, therefore,
  not self-consistent.  On the other hand, if the column density is lower
  ($N_{\rm OVII} \la 10^{17} ~\rm{cm}^{-2}$), the $f$ to $r$ ratio may be as
  low as $\sim 1$, in which case the total absorption $\ew$ is $\sim 35$ m\AA.
  The corresponding column density, in this case, is $N_{\rm OVII} \sim 2
  \times 10^{17} ~\rm{cm}^{-2}$, and is in much better agreement with our
  measurements.  Note that, for pure absorption by \ion{O}{7} at a column
  density of $\sim 3 \times 10^{18} ~\rm{cm}^{-2}$, the predicted He$\alpha$
  absorption $\ew$ is $\sim 50 - 60$ m\AA\ for $b = 100 - 130
  ~\rm{km~s}^{-1}$, which is much larger than observed in both the \chandra\
  \hetg\ and \xmm\ \rgs\ data.  We note that the above estimate for the amount
  of resonance absorption line re-filling is really an upper limit, since
  resonance line photons that are re-emitted or produced through recombination
  cascades also suffer subsequent scatterings as they travel through the
  photoionized medium.  For a more general and complete treatment of
  photoexcitation and photoionization, see \citet{ali02,ali03}.

  The estimate of the column density described above is fairly insensitive to
  the geometry of the absorbing medium, since we are comparing the $\ew$ of
  the forbidden line, which is produced entirely through recombination, to
  that of the resonance line.  There are, however, other possible geometric
  configurations that can, in principle, reproduce the observed \ion{O}{7}
  He-like triplet ratio.  We note that the problem with a high $\tau \sim 0.7$
  \ion{O}{7} column density interpretation is that the amount of resonance
  line re-filling as determined from the forbidden line flux is insufficient
  to account for the observed resonance absorption line $\ew$, as discussed
  above.  This problem can be avoided if one considers a region with low
  \ion{O}{7} column density (to enhance the resonance line relative to the
  forbidden line) and high covering fraction (to match the flux of the
  observed forbidden line).  We estimate that the required covering fraction
  must be approximately $10$ times larger than the covering of the absorber
  along the line of sight, and the column density must be $N_{\rm OVII} \sim
  10^{16} ~\rm{cm}^{-2}$.  In addition, this region cannot be intercepted by
  the absorber as seen from the observer, since the emitted resonance line
  photons will then be absorbed almost entirely.  Although this geometry is
  able to reproduce the observed \ion{O}{7} He$\alpha$ triplet ratio, it
  cannot reproduce the observed $\ew$ of the higher series lines.  Since the
  amount of re-filling in these lines is smaller compared to that of the
  He$\alpha$ line by roughly their respective oscillator strength ratio, it
  overpredicts the absorption $\ew$ substantially.  It is also rather ad hoc
  and contrived, as it would seem to make the identification of the absorber
  with the emitter and its self-consistency as argued above a pure
  coincidence.

  In addition, the high-velocity component does not seem to be present in the
  \chandra\ \hetg\ spectrum, possibly due to intrinsic variability of the
  line-of-sight warm absorber properties.  The \ion{O}{7} column density of
  this component, however, is more than an order of magnitude lower than that
  of component 1, and, therefore, imposes only a negligible contribution to
  the total \ion{O}{7} edge opacity.  On the other hand, the opacity of iron
  L-shell ions in this kinematic component, particularly those of \ion{Fe}{20}
  -- \ion{Fe}{23}, are non-negligible.  No comparison of the derived iron L
  properties with the \hetg\ data can be made, however, since \citet{lee01}
  did not publish the \chandra\ \hetg\ data above $E = 1 ~\rm{\kev}$ ($\lambda
  \la 12.4$~\AA).

  Variability of the warm absorber in the soft X-ray band ($0.5 \la E \la 2.5$
  \kev) was observed in \mcg\ by \asca\ \citep{fabian94,reynolds95,otani96}.
  It was suggested that the observed correlation of the \ion{O}{8} column
  density and the continuum flux, and the lack of variability in the
  \ion{O}{7} column density, are due to a two-zone nature of the absorbing
  medium.  \citet{orr97}, however, found that this model is not consistent
  with their \sax\ data.  \citet{morales00} later demonstrated that an
  additional, third zone can provide an adequate explanation.  As we have
  shown, however, the opacity above the \ion{O}{8} photoelectric edge is
  dominated by iron L ions.  The significance of the inferred correlation with
  flux, therefore, is unclear.

\subsubsection{\mrk}

\begin{figure}
  \includegraphics[height=6in,angle=-90]{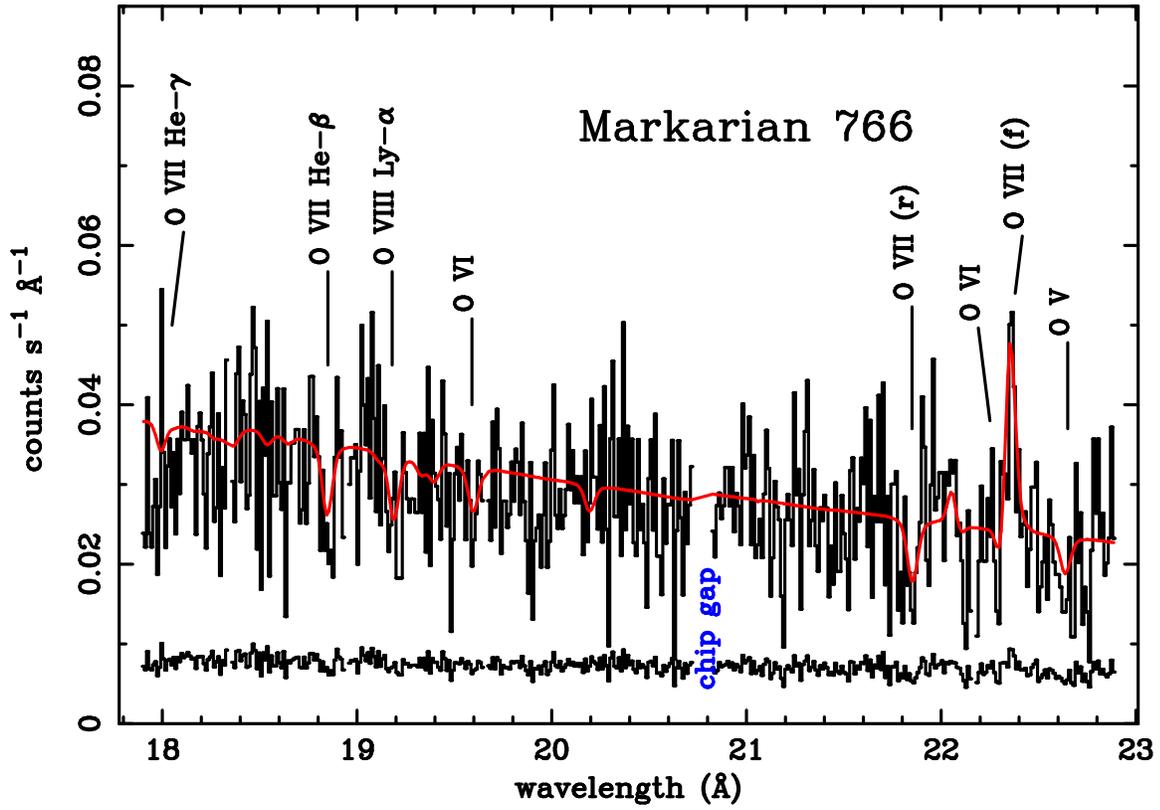}
  \caption{\mrk\ in the oxygen line region.  Absorption and emission lines
           from \ion{O}{8} -- \ion{O}{4} are clearly detected.  Only data
           from \rgs1 are shown.  The lower curve represents 1$\sigma$
           Poisson fluctuations.} \label{f6}
\end{figure}

  The absorption lines observed in \mrk\ are much less prominent than those in
  \mcg.  The spectrum, however, is of somewhat lower statistical quality,
  owing to the shorter exposure time and lower intrinsic flux during the
  observation (see Figure~\ref{f6}).

  In addition, multiple velocity components are not detected in the \mrk\
  spectrum.  Adopting a similar procedure for measuring the oxygen ion column
  densities as described above, we obtain $N_{\rm OV} = 6.1 \times 10^{16}
  ~\rm{cm}^{-2}$, $N_{\rm OVI} = 6.5 \times 10^{16} ~\rm{cm}^{-2}$, $N_{\rm
  OVII} = 8.9 \times 10^{16} ~\rm{cm}^{-2}$, and $N_{\rm OVIII} = 4.8 \times
  10^{16} ~\rm{cm}^{-2}$ (see Table~\ref{tbl3}).  One-parameter 90\%
  confidence ranges in the column densities are typically $\pm 0.5$ dex.  The
  best-fit turbulent velocity is $b = 90 ~\rm{km~s}^{-1}$.  These column
  densities are, again, much too low to produce observable photoelectric
  absorption.  No obvious velocity shifts are detected, with an upper limit of
  $|v| \sim 160 ~\rm{km~s}^{-1}$.

  The spectrum below $\lambda \la 17$~\AA\ is fairly well represented by a
  single powerlaw that matches the high-energy continuum derived from the
  \epic\ data \citep{boller01,page01}, and no obvious absorption features are
  detected from, for example, K-shell Ne and L-shell iron.  Upper limits to
  the derived column densities are also listed in Table~\ref{tbl3}.  As in the
  spectrum of \mcg, there is evidence for a broad absorption feature between
  $\lambda = 16 - 17$~\AA, which is most likely due to the presence of a weak
  iron \uta.

\subsection{Absorption by Dust Particles}
  \citet{lee01} argued that part of the sharp drop in the \mcg\ spectrum is
  due to absorption by iron oxides in dust grains.  The parameters derived by
  these authors (i.e., $N_{\rm Fe} = 3.5 \times 10^{17} ~\rm{cm}^{-2}$ and
  $N_{\rm O} = 7.0 \times 10^{17} ~\rm{cm}^{-2}$ in addition to Galactic
  absorption), however, are grossly incompatible with the \rgs\ data.  Using
  photoionization and resonance absorption cross sections of neutral iron
  calculated with \hullac, we measure the atomic column density to be $N_{\rm
  Fe} = (7 \pm 2) \times 10^{16} ~\rm{cm}^{-2}$, which is a factor of $\sim 5$
  lower than the derived \hetg\ value.  At least part of the reason for this
  discrepancy is due to the fact that the absorption cross sections published
  by \citet{kortright00}, which were used in the \hetg\ analysis, are
  underestimated owing to saturation effects through the sample filter.  The
  \rgs\ data do not show any obvious signatures of excess absorption by atomic
  oxygen at the systemic velocity of \mcg.  If we assume that the underlying
  continuum is smooth across the oxygen K-edge, the inferred column density is
  consistent with zero with a 90\% upper limit of $N_{\rm O} = 1 \times
  10^{16} ~\rm{cm}^{-2}$.  This is approximately two orders of magnitude lower
  than claimed for the \hetg\ \citet{lee01} model, probably again due to their
  limited number of photons in that region.  The shape of the spectrum across
  this wavelength region is also inconsistent with that of photoelectric
  absorption.  The spectrum, therefore, is inconsistent with any reasonable
  value for the Fe:O ratio in standard dust models (e.g., \citealt{snow96}).
  An Fe:O ratio of 1:2 with $N_{\rm Fe} = 3.5 \times 10^{17} ~\rm{cm}^{-2}$
  claimed by \citet{lee01} is fundamentally incompatible with the data (see
  Figure~\ref{f3}).  Even in the most optimistic case of an Fe:O ratio of 1:1,
  the neutral O edge is still severely overpredicted compared to what is
  observed in the spectrum.  Assuming an iron abundance of $A_{\rm Fe}/A_{\rm
  H} = 4.7 \times 10^{-5}$ \citep{anders89}, our measurement indicates an
  equivalent hydrogen column density of $N_{\rm H} = 1.5 \times 10^{21}
  ~\rm{cm}^{-2}$, which is a factor of few lower than the lower limit derived
  by \citet{reynolds97} from the optical reddening assuming the Galactic value
  for the dust-to-gas ratio.

  Laboratory measurements and theoretical calculations of
  \citet{crocombette95} first indicated that the iron L-shell opacity of iron
  oxides peaks near $\lambda \sim 17.75$~\AA, which is $\sim 200$ m\AA\ longer
  than what is observed in the \mcg\ spectrum.  However, recent communications
  with that group have now revealed that there was an error in the wavelength
  scale in that paper (Crocombette 2000, private communication;
  \citealt{gautier01,gota00}), and that the deepest resonance complex for iron
  oxides is instead centered at $17.46$~\AA.  While closer, this revised value
  still differs significantly from the wavelength of the observed edge-like
  feature in the \mcg\ spectrum, by more than the uncertainty attributable to
  the RGS instrument.  The observed position of this feature is much closer to
  that appropriate to pure iron, as measured by \citet{kortright00}, rather
  than iron oxides in dust grains.  The Fe L absorption feature observed in
  the spectrum must, therefore, be due to either atomic or pure metallic iron.

  Absorption by other elements that are expected to deplete onto dust are also
  not detected in the spectrum of \mcg.  For example, neutral Mg and Si, whose
  K-shell edges are at $\lambda = 9.48$~\AA\ and $\lambda = 6.74$~\AA,
  respectively, are not seen in the data.  The upper limits to their column
  densities are smaller than that of neutral oxygen with $N_{\rm Mg} \sim 4
  \times 10^{15} ~\rm{cm}^{-2}$ and $N_{\rm Si} \sim 8 \times 10^{15}
  ~\rm{cm}^{-2}$ at 90\% confidence.

  The spectrum of \mrk\ does not show any obvious evidence for absorption by
  neutral iron at the galaxy's systemic velocity.  Additional absorption by
  neutral oxygen, however, is possible.  If we assume that the underlying
  spectrum is smooth across the edge at $\lambda \sim 23$~\AA, the derived
  neutral oxygen column density is $N_{\rm O} = 1.5 \times 10^{17}
  ~\rm{cm}^{-2}$, but the model still leaves excess residuals just above the
  edge near $23.5 \la \lambda \la 26.0$~\AA.  In addition, the associated
  resonance $1s - 2p$ absorption line at $\lambda \sim 23.5$~\AA\
  \citep{krause94,stolte97,mclaughlin98,paerels01} is not detected at the
  galaxy's systemic velocity.  Resonance absorption by iron oxides in the
  range $\lambda \sim 23.3 - 23.4$~\AA\ \citep{wu97} is also not observed.
  The resulting empirical continuum is, again, flatter ($\Gamma \sim 1.4$) and
  brighter ($1.0 \times 10^{-2} ~\rm{photons~cm}^{-2} ~\rm{s}^{-1}
  ~\rm{\kev}^{-1}$ at 1 \kev) than that inferred from the high energy region.
  As for \mcg, the \mrk\ spectrum shows a complete absence of absorption
  features from other elements that are expected to deplete onto dust grains.

  The question still remains as to what causes the reddening in the optical
  and the UV band.  The \rgs\ spectrum of another highly-reddened \agn, \iras,
  does not show any obvious evidence for absorption by dust \citep{sako01}.
  The estimated reddening in this galaxy is $E(B-V) = 0.3$ \citep{wills92},
  which is $2 - 4$ times smaller than in \mcg, where $E(B-V) = 0.61 - 1.09$
  \citep{reynolds97} and comparable to that in \mrk\ ($E(B-V) = 0.4$;
  \citealt{walter93}).  Whether this is due simply to the variation in dust
  composition or an inherent inaccuracy of the $E(B-V)$ estimates is currently
  unknown.  It is also possible that the line of sight towards the X-ray
  source is nearly dust-free.

\section{The Relativistic Accretion Disk Interpretation}
\begin{figure}
  \includegraphics[height=4in,angle=-90]{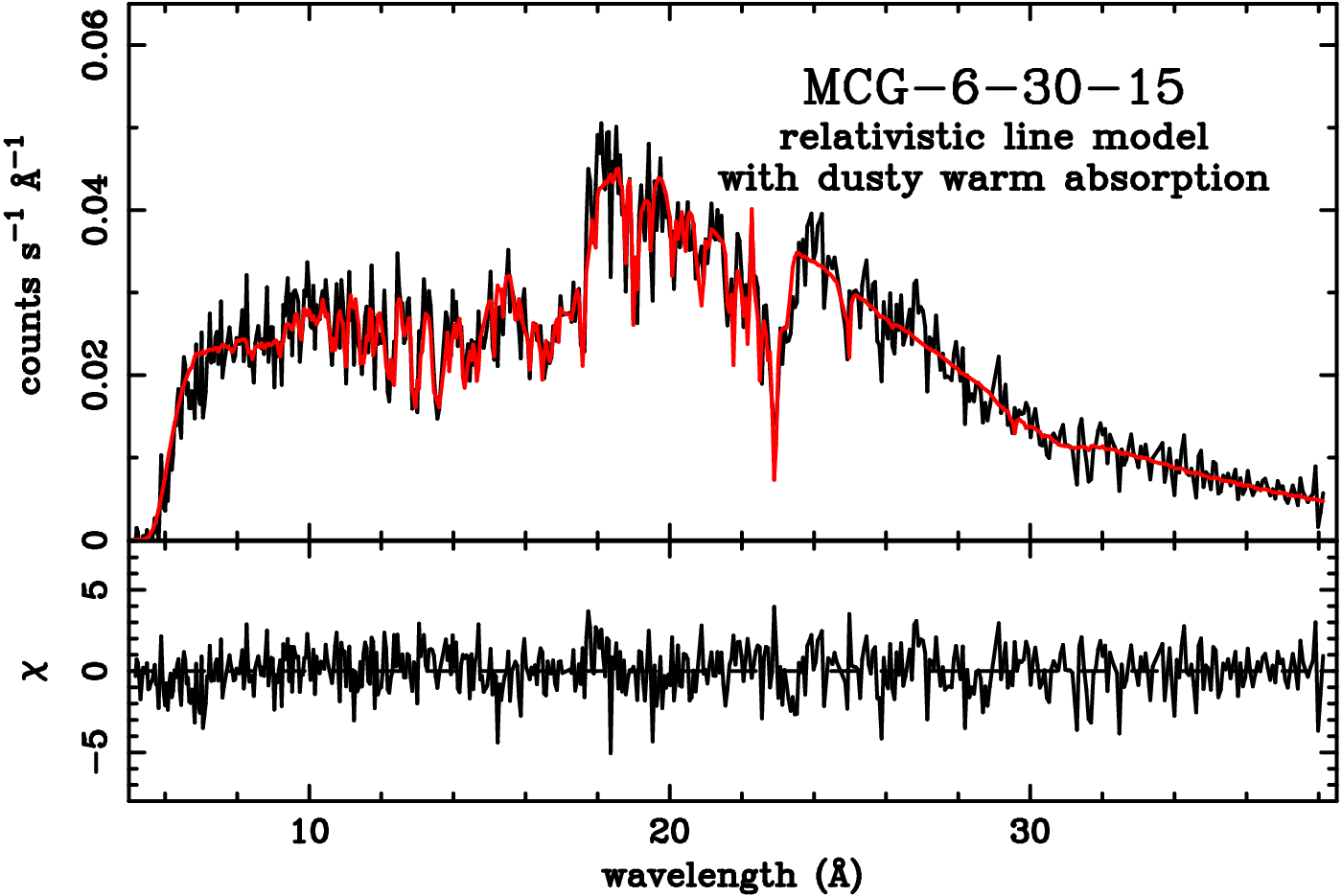}\\
  \includegraphics[height=4in,angle=-90]{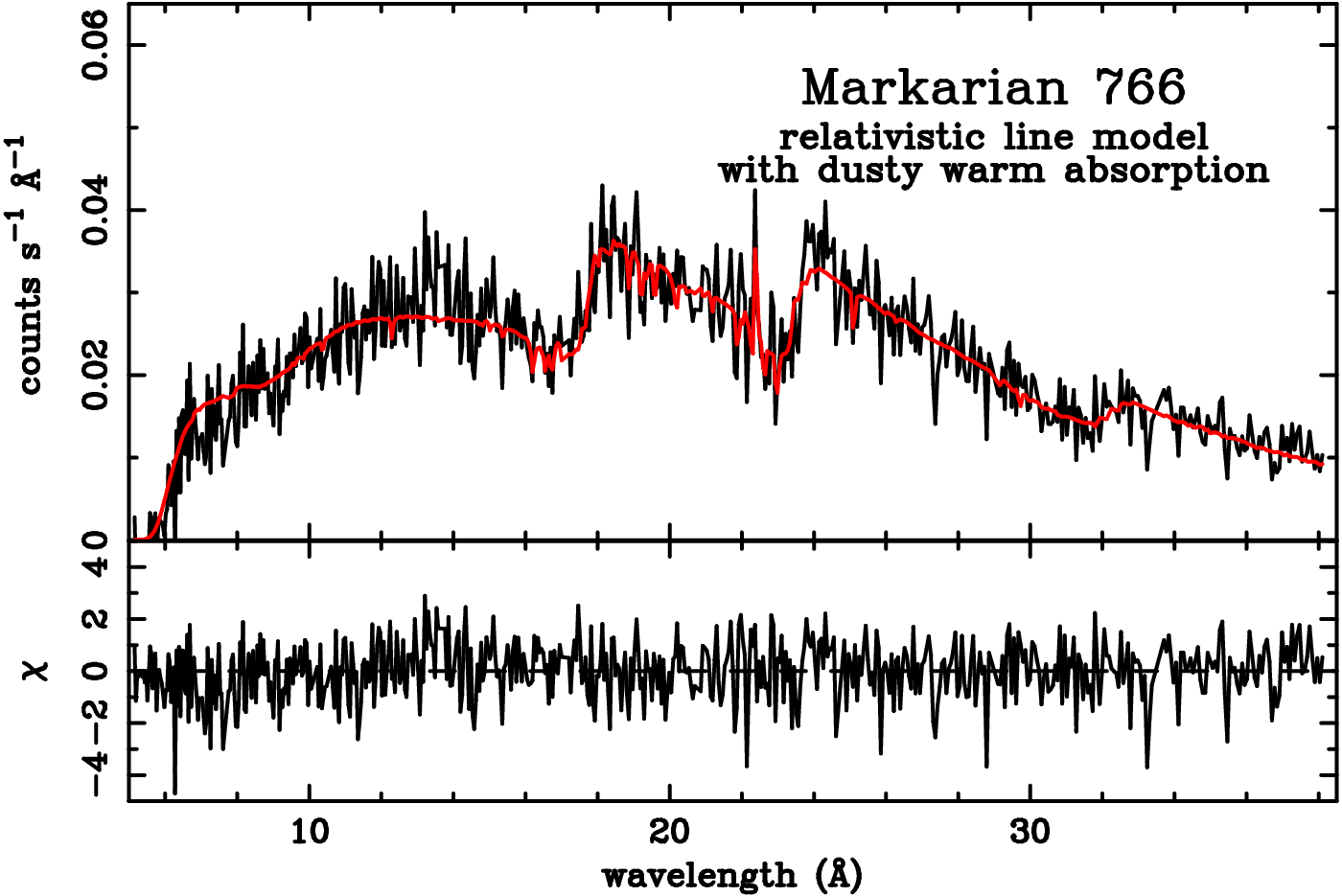}
  \caption{Best-fit relativistic line models for \mcg\ (top) and \mrk\
           (bottom).  No obvious residuals remain.}
           \label{f12}
\end{figure}

  As shown above, the ionic and atomic column densities measured from the fits
  to the absorption lines are too low to account for the spectral jumps
  observed in the data.  In the spectra of both \mcg\ and \mrk, the flux near
  $\lambda \sim 17.5$~\AA\ nearly doubles within a wavelength region of
  $\Delta \lambda \sim 0.3$~\AA.  We interpret these features as due to
  Ly$\alpha$ emission lines of carbon, nitrogen, and oxygen originating from
  the inner-most regions of a relativistic accretion disk, as in our original
  interpretation \citep{grazie01}.  The profiles of these lines appear to be
  different from those expected for a disk rotating in a Schwarzschild metric,
  which predicts prominent blue and red peaks (see e.g., \citealt{fabian89}).
  Instead, the profiles look similar to those produced in an accretion disk
  around a Kerr black hole, where the radius of marginal stability is
  substantially smaller than in a Schwarzschild metric, and, therefore, the
  effects of relativistic distortion are larger.  It is worth noting, however,
  that the nitrogen line in \mcg\ shows evidence for a red peak centered at
  $\lambda \sim 27$~\AA.  This feature is in a relatively clean band of the
  spectrum (i.e., no strong absorption lines are expected or observed), and
  contains $\sim 100 ~\rm{counts}$.  The best-fit global models for \mcg\ and
  \mrk\ using emission lines originating from an accretion disk around a Kerr
  metric \citep{laor91} are shown in Figure~\ref{f12}.  Statistically, this
  description of the data is better than that provided by the best-case dusty
  warm absorber model by $\Delta \chi^2 = 640$ for 1148 degrees of freedom.
  The best-fit accretion disk parameters are very similar to those derived in
  \citet{grazie01}, however, they are listed in Table~\ref{tbl4} for
  completeness.  Although we have chosen to tie the emissivity indices and the
  disk inner and outer radii for the three emission lines included in the fit,
  there is no reason to presume that these parameters for the three lines
  should be identical, as discussed briefly later in this section.

  Theoretical spectra originating from an X-ray ionized accretion disk have
  been investigated by many authors \citep{raymond93,ross93,matt93,ko94,
  zycki94}.  The details of the results predicted by the various models vary
  substantially, but there is one general property of an X-ray illuminated
  disk common to all model calculations -- the existence of a highly ionized
  layer (X-ray skin) on the surface of the disk with appreciable Thomson
  optical depth in the vertical direction.  This layer has a temperature close
  to the Compton temperature, which ranges from $T_C = ~\rm{few} \times 10^6 -
  10^8 ~\rm{K}$, depending on the shape of the ionizing continuum
  \citep{nayakshin00,nayakshin01}.  The medium is nearly fully ionized, with
  trace abundances of H-like ions of low- and mid-$Z$ elements, as well as the
  iron L species.

  As shown by \citet{nayakshin00} and \citet{nayakshin01}, the disk intrinsic
  flux has significant impact on the thermal/ionization structure of the
  surface layers of an accretion disk.  When the X-ray flux $F_X$ exceeds the
  disk intrinsic flux $F_d$, the ionization equilibrium curves (or S-curves)
  typically show three distinct thermally stable regions separated by unstable
  regions \citep{nayakshin01}.  However, if the disk flux is sufficiently
  large relative to the X-ray flux ($F_d \ga$ few $\times ~F_X$), the upper
  unstable region usually does not exist, mainly due to inverse Compton
  cooling by soft UV photons that affect the temperature structure at high
  ionization parameters.

\begin{figure}
  \includegraphics[height=6in,angle=-90]{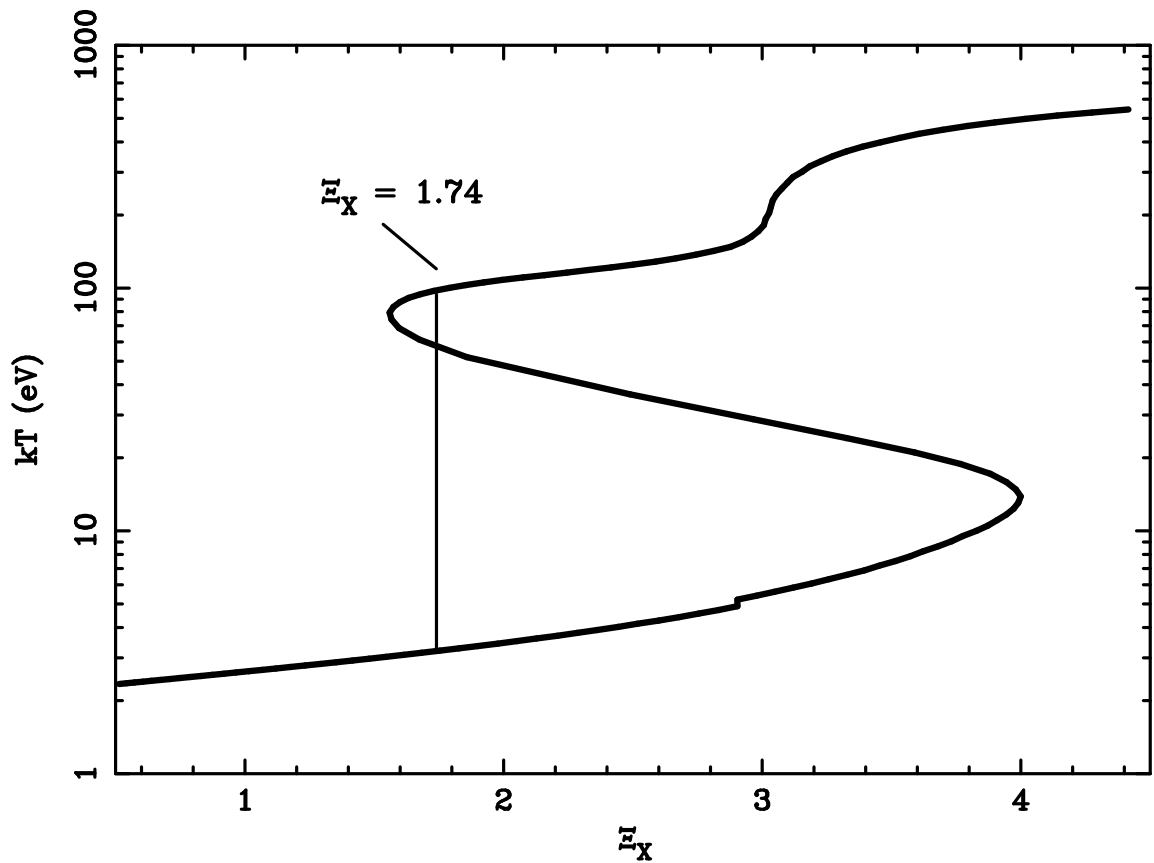}
  \caption{Ionization equilibrium curve (S-curve) for a $\Gamma = 1.8$
           powerlaw plus a $kT = 10 ~\rm{\ev}$ blackbody assuming an
           optically thin medium.  The ratio of the
           fluxes $F_{\rm{bb}}/F_{\rm{X}}$ is assumed to be 6 (see also
           \citealt{nayakshin01}).  The vertical line at $\Xi = 1.74$
           represents the ionization parameter at which the transition
           between the hot and cold phases occurs (see \citealt{li01}).}
           \label{f13}
\end{figure}

  To estimate the rough ionization and thermal structure of the irradiated
  disk, we use the photoionization code \xstar\ \citep{kallman95}, assuming
  that the medium is optically thin to the ionizing continuum.  The spectrum
  we adopt is a power law with $\Gamma = 1.8$ with a {\small UV} blackbody
  component of $kT = 10 ~\rm{\ev}$.  Although the UV flux of both \mcg\ and
  \mrk\ are unknown, we assume that the disk-to-X-ray flux ratio is $F_d/F_X =
  6$.  The resulting S-curve is shown in Figure~\ref{f13}.  We note that this
  value corresponds to the flux ratio as observed from the accretion disk
  surface and not from an observer at a distance.  The former may be larger by
  orders of magnitude, since the UV source lies just beneath the ionized
  surface layer, whereas the X-ray source is believed to lie above the disk at
  an unknown distance, which could be much larger than the distance to the UV
  source.  In order for our interpretation to hold, however, a value of
  $F_d/F_X \ga 6$ is required.

  \citet{nayakshin00} show that the Thomson optical depth in the hot layer is
  given approximately as, $\tau_h \sim (\Xi_t A)^{-1}$, where $\Xi_t$ is the
  pressure ionization parameter ($\Xi = F_{\rm X}/cP$, where $F_{\rm X}$ is
  the incident X-ray flux, $P$ is the pressure, and $c$ is the speed of light)
  at which the transition between the hot and cold phases occurs and $A$ is
  the dimensionless gravity parameter, which characterizes the ratio of the
  vertical component of the gravitational force at one disk scale height to
  the incident radiation force ($A = f_g/f_{\rm rad}$; see,
  \citealt{nayakshin01}).  Recently, \citet{li01} showed that thermal
  conduction between the hot and cold stable layers allows only a single,
  unique ionization parameter $\Xi_t$ where the transition can occur.
  Adopting their method, we find that $\Xi_t = 1.74$ for the ionizing
  continuum specified above.  The Thomson optical depth, therefore, is $\tau_h
  \sim 6~A_{0.1}^{-1}$, where $A_{0.1}$ is the gravity parameter in multiples
  of 0.1.  Although the numerical value for $A$ is uncertain, the derived
  Thomson depth is similar to our original estimate, which is based purely on
  the observed emission measures of the recombination lines \citep{grazie01}.

  Above the transition region at $\Xi_t \ga 1.74$, there are trace abundances
  of H-like ions of C, N, O, Ne, Mg, and Si, and substantial amounts of
  several charge states of L-shell Fe.  Ion fractions of He-like ions are
  almost completely negligible.  By integrating the fractional abundances
  along the hot stable branch, we find K-shell photoelectric absorption
  optical depths of H-like C, N, and O of $\tau = 1.3, 0.7, 9.8$,
  respectively, assuming solar abundances of \citet{anders89}.  A large
  fraction of the total photoelectric opacity is produced near the transition
  region at $1.74 \la \Xi \la 3$ where the fractional abundances of H-like C,
  N, and O are the highest.  This indicates that most of the soft X-ray flux
  is absorbed in this layer before penetrating into the lower branch of the
  S-curve.

  Since photoionization is balanced by recombination, essentially all of this
  flux is re-radiated into recombination line and continuum emission.  If the
  edge optical depth is large ($\tau \ga 1$), and the covering fraction of the
  illuminated disk is $\sim 2\pi$, the equivalent widths of the resulting line
  emission are, therefore, on the order of their respective ionization
  potentials.  For \ion{O}{8}, this amounts to approximately a few hundred \ev,
  much larger than what is observed in the \rgs\ data ($\ew \sim 160
  ~\rm{\ev}$).  Since the medium is almost completely optically thick to
  photoelectric absorption by \ion{O}{8}, any recombination emission at
  wavelengths shorter than that of the K-shell edge of \ion{O}{8} ($\lambda <
  14.22$~\AA), is absorbed and eventually re-radiated into \ion{O}{8}
  Ly$\alpha$, as suggested by \citet{grazie01}.  Recombination lines from Ne,
  Mg, Si, and most of the Fe L ions, lie above the \ion{O}{8} photoelectric
  edge, and, therefore, most of the line and continuum radiation are converted
  into \ion{O}{8} Ly$\alpha$.
  
  Higher order Lyman transitions of carbon, nitrogen, and oxygen should also
  be suppressed by line opacity, given the high photoelectric optical depths
  inferred.  After a few scatters, the excited levels for these lines decay
  via Balmer, Paschen, etc., emission, thereby contributing to the Ly$\alpha$
  lines.  However, a reliable estimate of the Lyman line opacity is difficult
  given the geometry of the problem.  One would naively expect the optical
  depth to be lowest in the vertical direction, but given the huge velocity
  shear in the disk, this is probably not the case.  A detailed radiative
  transfer model for this problem would require at least a 2D calculation,
  which is beyond the scope of this paper, and will be deferred to future
  publications.  So the absence of the higher Lyman series transitions in the
  observed spectra remains an open issue.

  In cases where the disk intrinsic flux is sufficiently low, the S-curve
  separates into three stable branches.  In such a case, the temperature in
  the hot layer is very close to the Compton temperature with only trace
  abundances of H-like C, N, and O.  The photoelectric opacities are,
  therefore, small, and the re-radiated spectrum is nearly featureless in the
  soft X-ray band.  This might explain why some Seyfert 1 galaxies exhibit
  these relativistic lines, and others clearly do not.

  Recently, \citet{ballantyne01} argued against the relativistic line
  interpretation of the soft X-ray features in \mcg\ based on their own model
  calculations of an irradiated disk.  They find that: (1) the predicted
  \ion{O}{8} Ly$\alpha$ line \ew\ is at most $\sim 10 ~\rm{\ev}$ (2)
  comparable emission line equivalent widths from \ion{O}{7} must be observed,
  and (3) that these lines must be accompanied by observable amounts of Fe L
  line emission (specifically, \ion{Fe}{17} and \ion{Fe}{18}) as well.  One of
  the reasons for the discrepancy with our conclusions above is the difference
  in the ionization structure of the hot layer.  For our calculations, the
  photoelectric opacity of \ion{O}{7}, \ion{Fe}{17}, \ion{Fe}{18} is $\tau
  \sim 0.2, 0.02, 0.1$, respectively, and the resulting line emission is
  negligible compared to that of \ion{O}{8}.  On the other hand, the opacities
  of \ion{Fe}{20} -- {\small XXVI} are non-negligible, and the recombination
  line fluxes are high enough to produce observable features, {\it if}
  radiative transfer effects are ignored.  As discussed above, however, the
  opacity of \ion{O}{8} is so large that almost all of the radiative power of
  \ion{Fe}{20} -- {\small XXVI} is re-radiated into \ion{O}{8} Ly$\alpha$.

  In a later publication, \citet{ballantyne02} were able to reproduce the
  large \ion{O}{8} $\ew$ inferred from the data, but criticized the
  relativistic line interpretation by showing that (1) the strength of the
  \ion{N}{7} cannot be reproduced and (2) Compton scattering smears out the
  blue edge of the disk line profile.  The former discrepancy can possibly be
  understood again as a radiative transfer effect, in which \ion{O}{8}
  Ly$\alpha$ line photons fluoresce with the \ion{N}{7} Ly$\zeta$ line in an
  optically thick medium, the details of which are discussed in
  \citep{sako03}.  In our view, the latter criticism is not a serious concern.
  Unlike in the case for iron, for example, where the abundance weighted
  K-shell photoelectric cross section is only a factor of few larger than the
  Thomson cross section, the abundance weighted K-shell photoelectric cross
  section of oxygen is more than two orders of magnitude larger for solar
  abundances.  Therefore, the amount of Compton scattering relative to
  \ion{O}{8} line production depends only on the trace abundance of \ion{O}{8}
  in their particular model assuming either constant disk density or
  hydrostatic equilibrium with a smooth density profile.  We speculate that
  reflection from an inhomogenous accretion disk, which more or less mixes
  regions from a range in ionization parameter, can suppress Compton
  scattering of the line by up to two orders of magnitude.  The viability of
  such models remains to be investigated quantitatively.  In any case, the
  discrepancy in the width of the blue edge is only a factor of few at most
  when the neutral iron \uta\ and edges are accounted for.

  \citet{vaughan01} noted that the lack of correlation of the iron K line flux
  with the continuum in \mcg\ cannot be understood in terms of a simple cold,
  irradiated disk, and suggested that this is likely to be due to the presence
  of an ionized layer above the disk.  Unless this layer is extremely hot,
  such that all the elements are completely ionized ($kT \ga ~\rm{few} \times
  10^7 ~\rm{K}$), recombination emission must be produced and should be
  observable, if the opacity is substantially high.

  \citet{wilms01} have shown that the \epic\ \pn\ data of \mcg\ exhibit an
  extremely broad line at a rest energy of $E \sim 6.97 ~\rm{\kev}$, which is
  most likely produced by H-like iron.  For the model described above, the
  photoelectric opacity of \ion{Fe}{26} through the disk is $\tau \sim 0.7$,
  while that of \ion{Fe}{25} is $\tau \sim 1.0$.  Therefore, a significant
  fraction of the incident continuum above $E \sim 9 ~\rm{\kev}$ is re-radiated
  in lines, and the observed equivalent width of $EW \sim 300 - 400 ~\rm{\ev}$
  can be qualitatively understood.  The radial emissivity index inferred from
  the Fe line, however, is slightly larger than that inferred from the \rgs\
  data using the C, N, and O lines, but within the error bars.  We note that
  the derived indices, in general, do not need to be identical, since, if the
  lines are produced through recombination, the radial emissivity profile does
  not necessarily trace the density profile.  Various effects, such as
  temperature and ion fraction gradients, may be able to explain the
  discrepancy.  Also, as noted by \citet{wilms01}, returning radiation from
  the disk may affect the observed emissivity profile as well.  A more
  quantitative investigation is certainly required, and will be discussed in a
  future publication.

  If there are fluorescent iron lines produced below the ionized skin in a
  cold medium (in the cold stable branch of the S-curve), they must suffer
  multiple Compton scatterings as they travel through the hot medium.  Since
  the temperature of the hot layer is still substantially lower than the iron
  K line energy, the photons will be down-scattered and broadened in addition
  to being subjected to relativistic effects (see also, \citealt{misra99}).
  Note that this hot layer does not alter the shape of the direct continuum
  radiation, which is presumably formed above the accretion disk (cf.,
  \citealt{reynolds00, ruszkowski00}).  We also note that most of the iron
  L-shell ions (\ion{Fe}{18} -- {\small XXIV}) that produce the forest of
  valence-shell absorption lines observed in the \rgs\ data also absorb the
  $6.4 \la E \la 6.7$ \kev\ region in the inner-shell resonance transitions.
  Using the measured column densities of L-shell iron ions listed in
  Table~\ref{tbl2} for \mcg, we estimate the total absorption equivalent width
  to be $\ew \sim 30$ \ev, capable of carving out a non-negligible portion of
  the blue emission edge of the relativistic line profile.  Accounting for
  this opacity is important for detailed analyses of the broad iron K lines in
  Seyfert 1 galaxies.

\section{Summary}

  By taking advantage of the extended wavelength coverage and high statistical
  quality provided by \xmm, we have shown that the dusty warm absorber model
  posited by \citet{lee01} cannot explain the measured \rgs\ spectra of either
  \mcg\ or \mrk.  The ionized oxygen column densities and turbulent velocity
  widths are well-constrained by the data, and eliminate the possibility that
  ionized oxygen makes any significant contribution to the observed edge-like
  feature near 17.5~\AA.  The absence of any discernible neutral oxygen edge,
  and the detailed structure of the 17.5~\AA\ feature further suggest that
  absorption by iron oxides associated with dust is negligible in both
  spectra.  By contrast, the model involving relativistic Ly$\alpha$ emission
  lines of carbon, nitrogen, and oxygen, with some contribution from an
  ionized absorber, quantitatively provides a superior fit.  The parameters
  characterizing these emission lines are roughly compatible with what would
  be expected for the surface layers of an irradiated accretion disk.  While
  there are still outstanding theoretical questions associated with this
  interpretation, the relativistic line model more successfully reproduces the
  spectra of these two objects.

\acknowledgements We thank the anonymous referee for constructive comments,
  which helped us improve the clarity and the presentation of the paper.  This
  work is based on observations obtained with \xmm, an \esa\ science mission
  with instruments and contributions directly funded by \esa\ Member States
  and the {\small USA} (\nasa).The Columbia University team is supported by
  \nasa\ through the \xmm\ mission support and data analysis.  {\small MS} was
  partially supported by \nasa\ through \chandra\ Postdoctoral Fellowship
  Award Number {\small PF}1-20016 issued by the \chandra\ X-ray Observatory
  Center, which is operated by the Smithsonian Astrophysical Observatory for
  and behalf of \nasa\ under contract {\small NAS}8-39073.  The Mullard Space
  Science Laboratory acknowledges financial support from the UK Particle
  Physics and Astronomy Research Council.  The Laboratory for Space Research
  Utrecht is supported financially by the Netherlands Organization for
  Scientific Research ({\small NWO}).  Work at {\small LLNL} was performed
  under the auspices of the U. S. Deparment of Energy by the University of
  California Lawrence Livermore National Laboratory under contract
  No. W-7405-Eng-48.

\clearpage

\begin{deluxetable}{cccrcccr}
  \tabletypesize{\scriptsize}
  \tablewidth{0pt}
  \tablecaption{\xmm\ Observation Log \label{tbl1}}
  \tablehead{
    \colhead{} & \multicolumn{3}{c}{\mcg} & \colhead{} &
      \multicolumn{3}{c}{\mrk} \\
    \cline{2-4} \cline{6-8} \\
    \colhead{Instrument} & \colhead{Mode} & \colhead{Filter} &
      \colhead{Exposure (s)\tablenotemark{a}} & & \colhead{Mode} &
      \colhead{Filter}\tablenotemark{a} &
      \colhead{Exposure (s)\tablenotemark{b}}
  }
  \startdata
    RGS1 & Spectroscopy + Q & \nodata & 127272 & & Spectroscopy + Q &
            \nodata & 56593 \\
    RGS2 & Spectroscopy + Q & \nodata & 127543 & & Spectroscopy + Q &
            \nodata & 56613 \\
    MOS1 & Timing -- uncompressed & MEDIUM & 82753 & & 
           Imaging -- full window & MEDIUM & 28602 \\
    MOS2 & Imaging -- full window & MEDIUM & 93904 & & 
           Imaging -- full window & MEDIUM & 28280 \\
    PN   & Imaging -- small window & MEDIUM & 69662 & & 
           Imaging -- small window& MEDIUM & 37127 \\
    OM   & Imaging & UVW2   & 82328 & & Imaging & UVW1 & 10000 \\
    \nodata   & \nodata & \nodata   & 82328 & & Imaging & UVM2 & 19500 \\
    \nodata   & \nodata & \nodata   & \nodata & & Imaging & UVW2 & 9600 \\
  \enddata
  \tablenotetext{a}{where applicable}
  \tablenotetext{b}{total good time}
\end{deluxetable}

\clearpage

\begin{deluxetable}{lrr}
  \tabletypesize{\footnotesize}
  \tablewidth{0pt}
  \tablecaption{Derived Absorption Column Densities for \mcg \label{tbl2}}
  \tablehead{
    \colhead{} & \multicolumn{2}{c}{$\log N_i$ (cm$^{-2}$)} \\
    \cline{2-3} \\
    \colhead{Ion} & \colhead{Component 1} & \colhead{Component 2}
  }
  \startdata
    \ion{C}{6}   & $\leq$ 16.43\tablenotemark{a} & $\leq$ 16.29\tablenotemark{a} \\
    \ion{N}{5}   & 15.52 & $\leq$ 16.58\tablenotemark{a} \\
    \ion{N}{6}   & 15.01 & $\leq$ 16.87\tablenotemark{a} \\
    \ion{N}{7}   & 16.57 & 16.11 \\
    \ion{O}{5}   & 16.94 & $\leq$ 16.23\tablenotemark{a} \\
    \ion{O}{6}   & 16.55 & 15.98 \\
    \ion{O}{7}   & 17.34 & 16.13 \\
    \ion{O}{8}   & 17.27 & 16.75 \\
    \ion{Ne}{9}  & 16.54 & 16.08 \\
    \ion{Ne}{10} & 17.00 & 16.83 \\
    \ion{Mg}{11} & 16.12 & 16.14 \\
    \ion{Mg}{12} & $\leq$ 16.74\tablenotemark{a} & 16.73 \\
    \ion{Si}{13} & $\leq$ 17.10\tablenotemark{a} & 16.52 \\
    \ion{Si}{14} & $\leq$ 17.03\tablenotemark{a} & $\leq$ 17.54\tablenotemark{a} \\
    \ion{Fe}{17} & 16.42 & 15.72 \\
    \ion{Fe}{18} & 16.97 & 16.37 \\
    \ion{Fe}{19} & 17.02 & 16.37 \\
    \ion{Fe}{20} & 16.49 & 16.81 \\
    \ion{Fe}{21} & 16.73 & 16.95 \\
    \ion{Fe}{22} & 16.66 & 16.69 \\
    \ion{Fe}{23} & 15.07 & 17.05 \\
    \ion{Fe}{24} & $\leq$ 16.42\tablenotemark{a} & 16.61 \\
  \enddata
\tablenotetext{a}{90\% upper limit.}
\end{deluxetable}

\clearpage

\begin{deluxetable}{lrr}
  \tabletypesize{\footnotesize}
  \tablewidth{0pt}
  \tablecaption{Derived Absorption Column Densities for \mrk \label{tbl3}}
  \tablehead{
    \colhead{Ion} & \colhead{$\log N_i$ (cm$^{-2}$)} \\
  }
  \startdata
    \ion{C}{6}   & $\leq$ 16.84\tablenotemark{a} \\
    \ion{N}{5}   & $\leq$ 16.74\tablenotemark{a} \\
    \ion{N}{6}   & 16.28 \\
    \ion{N}{7}   & 16.33 \\
    \ion{O}{5}   & 16.78 \\
    \ion{O}{6}   & 16.81 \\
    \ion{O}{7}   & 16.69 \\
    \ion{O}{8}   & 16.95 \\
    \ion{Ne}{9}  & $\leq$ 17.24\tablenotemark{a} \\
    \ion{Ne}{10} & $\leq$ 17.60\tablenotemark{a} \\
    \ion{Mg}{11} & $\leq$ 17.12\tablenotemark{a} \\
    \ion{Mg}{12} & $\leq$ 16.50\tablenotemark{a} \\
    \ion{Si}{13} & $\leq$ 17.11\tablenotemark{a} \\
    \ion{Si}{14} & $\leq$ 17.53\tablenotemark{a} \\
    \ion{Fe}{17} & $\leq$ 17.01\tablenotemark{a} \\
    \ion{Fe}{18} & $\leq$ 16.88\tablenotemark{a} \\
    \ion{Fe}{19} & $\leq$ 16.58\tablenotemark{a} \\
    \ion{Fe}{20} & $\leq$ 16.26\tablenotemark{a} \\
    \ion{Fe}{21} & $\leq$ 16.96\tablenotemark{a} \\
    \ion{Fe}{22} & $\leq$ 16.39\tablenotemark{a} \\
    \ion{Fe}{23} & $\leq$ 17.16\tablenotemark{a} \\
    \ion{Fe}{24} & $\leq$ 17.33\tablenotemark{a} \\
  \enddata
\tablenotetext{a}{90\% upper limit.}
\end{deluxetable}

\clearpage

\begin{deluxetable}{lrr}
  \tabletypesize{\footnotesize}
  \tablewidth{0pt}
  \tablecaption{Derived Accretion Disk Parameters \label{tbl4}}
  \tablehead{
    \colhead{} & \multicolumn{2}{c}{Value} \\
    \cline{2-3} \\
    \colhead{Parameter} & \colhead{\mcg} & \colhead{\mrk}
  }
  \startdata
  $i$ (degrees) & $38.5 \pm 0.4$ & $34.3 \pm 0.74$ \\
  $q$ & $4.49 \pm 0.15$ & $3.66 \pm 0.22$ \\
  $R_{\rm in}$ ($GM/c^2$) & $3.21 \pm 1.2$ & $1.25^{+1.1}_{-0.0}$ \\
  $R_{\rm out}$ ($GM/c^2$) & $100^{+95}_{-48}$ & $80.6^{+73}_{-20}$ \\
  $EW_{\rm C~VI}$ & $24.9 \pm 2.5$ & $28.3 \pm 2.9$ \\
  $EW_{\rm N~VII}$ & $54.4 \pm 3.2$ & $72.3 \pm 3.8$ \\
  $EW_{\rm O~VIII}$ & $162 \pm 8$ & $141 \pm 9$ \\
  \enddata
\end{deluxetable}

\clearpage

\begin{figure}
  \includegraphics[height=6in,angle=-90]{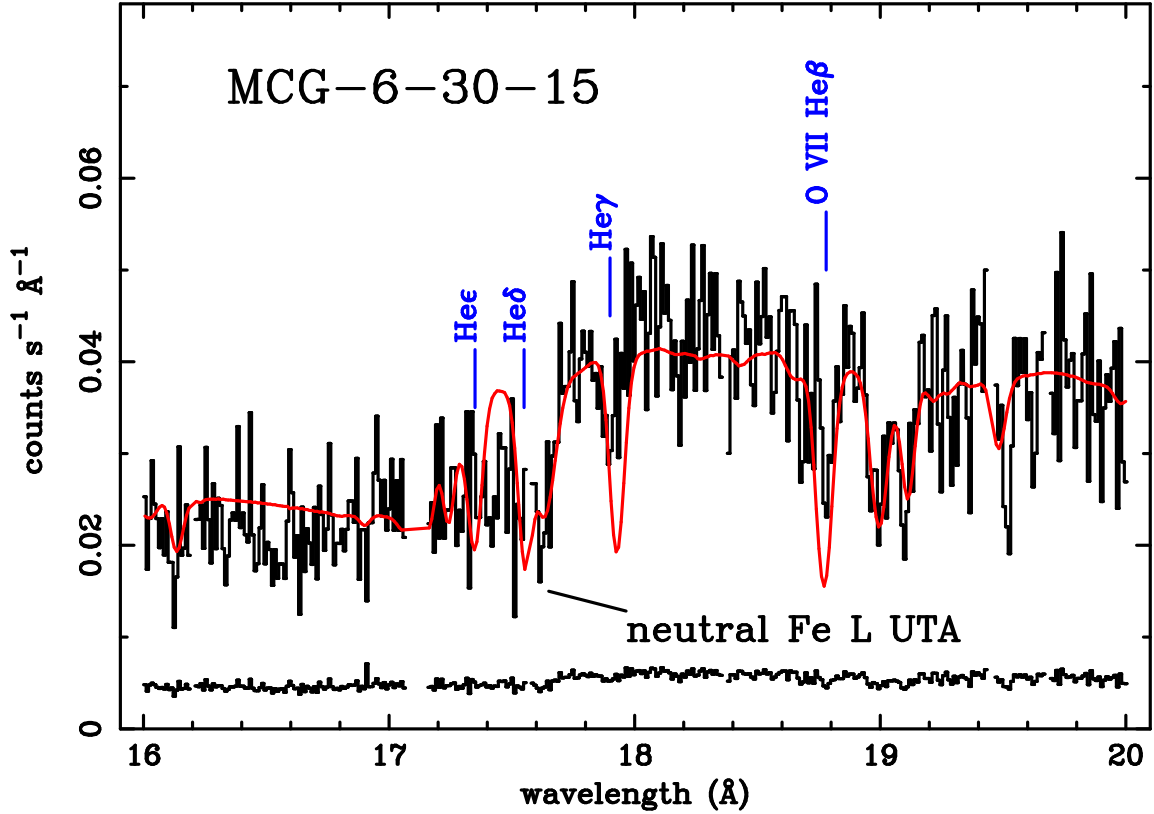}
  \caption{The $16 - 20$~\AA\ region of the \mcg\ spectrum showing the higher
           series lines of \ion{O}{7} and the neutral Fe L \uta\ using the
           \citet{lee01} dusty warm absorber model, with a reduced neutral Fe
           L column density ($N_{\rm Fe} = 7 \times 10^{16} ~\rm{cm}^{-2}$).
           The absorption line equivalent widths of \ion{O}{7} are severely
           overpredicted compared to the data.  In addition, the model flux
           between the He$\delta$ and He$\epsilon$ lines is overpredicted
           owing to the lack of \ion{O}{7} and Fe L opacity in this spectral
           region.}
           \label{f11}
\end{figure}

\end{document}